\documentclass{article}
\usepackage{graphicx,fancyhdr,tabularx,amsmath,amssymb,color}
\usepackage{subcaption}
\usepackage[utf8]{inputenc}
\usepackage[english]{babel}

\newlength{\threesubht}
\newsavebox{\threesubbox}

\usepackage{subcaption}
\usepackage{physics}
\usepackage{authblk}
\usepackage{breqn}
\usepackage[export]{adjustbox}
\usepackage{url}
\usepackage{float}
\usepackage{svg}
\usepackage{titlesec, blindtext, color}
\definecolor{gray75}{gray}{0.75}
\usepackage{verbatim}
\usepackage{multirow}
\usepackage{titlesec}
\usepackage{todonotes}

\NewDocumentCommand{\evalat}{sO{\big}mm}{%
  \IfBooleanTF{#1}
   {\mleft. #3 \mright|_{#4}}
   {#3#2|_{#4}}%
}

\usepackage{listings}
\usepackage{xcolor}
 
\definecolor{codegreen}{rgb}{0.2,0.7,0.2}
\definecolor{codeblue}{rgb}{0,0.2,0.5}
\definecolor{codered}{rgb}{0.7,0.3,0.3}
\definecolor{backcolour}{rgb}{0.95,0.95,0.92}
\lstdefinestyle{mystyle}{
    backgroundcolor=\color{backcolour},   
    commentstyle=\color{codeblue},
    keywordstyle=\color{codegreen},
    stringstyle=\color{codered},
    basicstyle=\ttfamily\footnotesize,
    breakatwhitespace=false,         
    breaklines=true,                 
    captionpos=b,                    
    keepspaces=true,                 
    numbers=left,                    
    numbersep=5pt,                  
    showspaces=false,                
    showstringspaces=false,
    showtabs=false,                  
    tabsize=2
}
\usepackage[
backend=biber,
sorting=none,
defernumbers=true,
giveninits=true,
maxbibnames=10,
doi=false,
isbn=false,
url=false,
eprint=false,
arxiv=false,
natbib=true]{biblatex}
\makeatletter
\let\blx@rerun@biber\relax
\makeatother

\newcommand{\vv}[1]{\boldsymbol{#1}}

\newcommand{\fr}{\mbox{$\frac{1}{2}$}}

\usepackage[a4paper, total={6.3in, 9in}]{geometry}

\title{A numerical method for the simulation of viscoelastic fluid surfaces}
\author[1]{Eloy de Kinkelder}
\author[2]{Leonard Sagis}
\author[1,3]{Sebastian Aland}
\affil[1]{HTW Dresden - University of Applied Sciences, 01069 Dresden, Germany}
\affil[2]{Wageningen University \& Research, 6708 PB Wageningen, the Netherlands}
\affil[3]{TU Bergakademie Freiberg, 09599 Freiberg, Germany}

\date{\today}

\begin{document}

\maketitle

\begin{abstract}
Viscoelastic surface rheology plays an important role in multiphase systems. A typical example is the actin cortex which surrounds most animal cells.
It shows elastic properties for short time scales and behaves viscous for longer time scales. Hence, realistic simulations of cell shape dynamics require a model capturing the entire elastic to viscous spectrum.
However, currently there are no numerical methods to simulate deforming viscoelastic surfaces. Thus models for the cell cortex, or other viscoelastic surfaces, are usually based on assumptions or simplifications which limit their applicability. 
In this paper we develop a first numerical approach for simulation of deforming viscoelastic surfaces. 
To this end, we derive the surface equivalent of the upper convected Maxwell model using the GENERIC formulation of nonequilibrium thermodynamics. The model  distinguishes between shear dynamics and dilatational surface dynamics. The viscoelastic surface is embedded in a viscous fluid modelled by the Navier-Stokes equation. Both systems are solved using Finite Elements. The fluid and surface are combined using an Arbitrary Lagrange-Eulerian (ALE) Method that conserves the surface grid spacing during rotations and translations of the surface. We verify this numerical implementation against analytic solutions and find good agreement. To demonstrate its potential we simulate the experimentally observed tumbling and tank-treading of vesicles in shear flow. We also supply a phase-diagram to demonstrate the influence of the viscoelastic parameters on the behaviour of a vesicle in shear flow. Finally, we explore cytokinesis as a future application of the numerical method by simulating the start of cytokinesis using a spatially dependent function for the surface tension. 
\end{abstract}

\section{Introduction}
Surface rheology plays a dominant role in multiphase systems such as foams, thin films, membranes, emulsions and polymer blends \cite{brenner2013interfacial,slattery2007interfacial}. It can affect the rate of coalescence in foams and emulsions, the deformation of droplets, vesicles or cells in flow, or the rise velocity of bubbles and emulsion droplets in a stagnant fluid.

Surface rheological properties are divided into in-plane (surface dilation and shear) and out-of-plane (surface bending) contributions. While out-of-plane viscoelastic behavior has been rigorously modeled \cite{rey2006polar}, 
in-plane viscoelasticity is mostly analyzed with simplified models \cite{sagis2010modelling} which generalize bulk behavior to the surface in an ad-hoc manner, neglecting the subtleties of the rheology on a surface. 
One of these subtleties is the presence of a finite dilational modulus of the surface fluid which is fundamentally different from bulk models which are designed for incompressible fluids. Another issue is the role of curvature leading to special surface tensorial derivatives \cite{nitschke_surface_time_derivatives}. 

One crucial example for in-plane viscoelastic surface rheology is the cell cortex. This thin ($<0.5\,\mu$m), cross-linked network of the polymerized protein actin underpins the membrane surrounding animal cells. 
The cortex is a key regulator for the emergence of cell shape and vital for cell function, for instance during cell division. 
On short time scales ($<1$ s) the cortex exhibits a dominantly elastic in-plane response to external mechanical stresses. It is the main determinant of stiffness of the cell surface in mammalian cells \cite{Salbreux2012536,FISCHERFRIEDRICH2016589}.
On larger time scales ($>10$ s) the cortex can undergo dynamic remodeling because of network rearrangement which results in a fluid-like rheology \cite{FISCHERFRIEDRICH2016589}. 
There is increasing evidence that this dynamic plasticity is an important prerequisite for cell viability in a changing extracellular environment, because it allows to rapidly change shape, move and exert forces \cite{Salbreux2012536}. 
Despite the overwhelming importance of the cortex in cell biomechanics, the lack of a numerical method for a viscoelastic surface implies that numerical results of cortex dynamics are currently limited to either purely viscous \cite{Mietke2019} or purely elastic behavior \cite{mokbel2020poisson}.

Most numerical methods that couple surface rheology to surrounding fluids are designed for membranes and red blood cells. These models include bending stiffness and describe the surface either as an inextensible membrane (infinite dilational modulus) \cite{ong2020immersed,veerapaneni2009boundary,aland2014diffuse,Marth2015,lowengrub2016numerical}
or as an elastic shell (with finite shear and dilational modulus) \cite{pozrikidis2010computational, le2009implicit, mokbel2020ale,Mokbel2017}.
Only in the last decade the first numerical models were proposed to simulate fluidic surfaces, yet assuming purely viscous behavior \cite{reuther2018solving, reuther2016incompressible, jankuhn2019higher, reusken2020stream, ArroyoDeSimone_PRE_2009}. 
A combination of both, namely a numerical method for a \textit{viscoelastic}, deforming fluid surface has not been considered. 

For bulk fluids the standard formulation of viscoelasticity is given by the upper-convected Maxwell model, in which dissipation is proportional to elastic shear stresses. Simple variations lead to more advanced formulations, like the Giesekus or Phan-Thien-Tanner model \cite{Phan-Thien2013}.
A first attempt to formulate a surface equivalent of the upper-convected Maxwell model was taken in \cite{sagis2010modelling,sagis2011dynamic}.  In this model both surface shear and dilatational effects were accounted for by formulating separate constitutive equations for the deviatoric part and the trace of the surface stress tensor. However that particular form of the upper-convected derivative used in the expression for the deviatoric stress does not conserve the zero trace of that tensor, which means this model can be applied only for small deformation rates and short time scales. 


In this paper, we present a first numerical method for the simulation of a viscoelastic surface. 
To this end, we first present a modified version of the model from \cite{sagis2010modelling,sagis2011dynamic} which fixes the problems mentioned above. The new model is rigorously derived 
using the GENERIC formulation of non-equilibrium thermodynamics. 
We show that for a flat and stationary surface, the model reduces to an ordinary two-dimensional Maxwell fluid model. 
As real surfaces are always surrounded by bulk phases, we embed the surface into viscous fluids governed by the Navier-Stokes equations. The complete system is discretized with a special Arbitrary Lagrangian-Eulerian (ALE) method, where grid movement is constructed such as to obtain a proper mesh despite surface translations and rotations. 
The method can exactly represent the discontinuous pressure at the interface by using augmented Taylor-Hood finite-element spaces at the surface. 
In a set of numerical tests we validate the method against analytical solutions for the shear and dilatational part separately. 
Finally, we illustrate the potential of the method by providing the first simulation of a viscoelastic fluid surface in shear flow. Also, we provide a phase diagram which nicely illustrates how viscoelastic parameters determine the transition from tank-treading to tumbling behavior. 

The rest of the paper is structured as follows. In Sec.~\ref{sec: surface stress derivation} the stress of a viscoelastic surface is derived.
The surface rheology is coupled to surrounding fluids in Sec.~\ref{sec:Model}. The numerical discretization by a finite element formulation of the coupled bulk/surface problem is presented in Sec.~\ref{sec:Discretization}. The numerical results are compared with analytical solutions, separately for the dilational and shear components, in Sec.~\ref{sec:num}.
Finally, the first numerical simulation results of viscoelastic surfaces in viscous fluids are provided by simulating a vesicle in shear flow and simulating the onset of cytokines. 
Conclusions are drawn in Sec.~\ref{sec:conclusion}.

\section{Stress of a viscoelastic surface}
\label{sec: surface stress derivation}
In this section we will use the GENERIC formulation of non-equilibrium thermodynamics \cite{Grmela1997,Oettinger1997,Oettinger2005} to derive the upper-convected Maxwell model for the surface stress associated with a viscoelastic interface. In the GENERIC formulation the dynamics of a multiphase system are described by a single equation of the form
\begin{equation} \label{Generic:definition}
\frac{d A}{dt} =\{A,E\}+\left[A,{\cal S}\right],
\end{equation}
where $A$ is an arbitrary observable of the system. The Poisson bracket $\{A,E\}$ describes the reversible part of the dynamics, and the dissipative bracket $\left[A,S\right]$ describes the irreversible part. Here $E$ is the total energy of the system and ${\cal S}$ is its total entropy. The specific form of the Poisson and dissipative brackets depends on the set of bulk and surface system variables, and is restricted by several conditions: the Poisson bracket is anti-symmetric ($\{A,B\}=-\{B,A\}$) and satisfies the Jacobi identity; the dissipative bracket is symmetric and satisfies $\left[A,A\right]\geq 0$. These brackets also satisfy the degeneracy conditions $\{A,{\cal S}\}=0$ and $\left[A,E\right]=0$  \cite{Grmela1997,Oettinger1997,Oettinger2005}.

For the sake of simplicity we will focus here only on the contributions to the Poisson and dissipative brackets from a single surface tensor $M$ and scalar variable $\Lambda$, i.e. we assume that our arbitrary observable $A$ can be expressed as
\begin{equation}\label{eq 12}
A=\int_{\Gamma} a(M, \Lambda )d\vv{x},
\end{equation}
where $a(M, \Lambda )$ is the density of A on the interface $\Gamma$. The dynamics of the system with respect to this restricted variable space is governed by the equation
\begin{equation} \label{eq 13}
\frac{d A}{dt} =\{A,E\}_{M,\Lambda}+\left[A,S\right]_{M,\Lambda},
\end{equation}
where $E$ and  ${\cal S}$ are given by
\begin{equation}
E=\int_\Gamma \left( \frac{\vv{m}^2}{2\rho_\Gamma} + u_\Gamma(M,\Lambda)\right)d\vv{x} \ \ \ \ \  {\cal S}= \int_\Gamma s_\Gamma\left( \rho_\Gamma, s_\Gamma(M,\Lambda)\right)d\vv{x}
\end{equation}
and the brackets $\{A,E\}_{M,\Lambda}$ and $\left[A,S\right]_{M,\Lambda}$ are the parts of the Poisson and dissipative brackets associated with $M$ and $\Lambda$. $\vv{m}$ is the surface momentum density (=$\rho_\Gamma\vv{v}_\Gamma$, where $\rho_\Gamma$ is the surface density and $\vv{v}_\Gamma$ is the surface velocity). $u_\Gamma$ is the surface internal energy density, and $s_\Gamma$ the surface entropy density. 
Let $C$ be an arbitrary symmetric surface tensor, with nonzero trace.  The Poisson bracket $\left\{A,B\right\}_C$ (where $B$ is arbitrary) for this tensor is given by \cite{Oettinger2005,Sagis-Oettinger2013}
\begin{equation} \label{eq 15}
\begin{split}
\left\{A,B\right\}_C = & \int_\Gamma C_{jk}\nabla_{\Gamma l} \left(\frac{\delta A}{\delta C_{jk}} \frac{\delta B}{\delta m_{l}} -  \frac{\delta B}{\delta C_{jk} }\frac{\delta A}{\delta m_{l}} \right)d\vv{x} \\ 
& +  \int_\Gamma  C_{lk} \left(\frac{\delta A}{\delta C_{jk}}\nabla_{\Gamma l} \frac{\delta B}{\delta m_{j}} -  \frac{\delta B}{\delta C_{jk} }\nabla_{\Gamma l}\frac{\delta A}{\delta m_{j}} \right)d\vv{x} \\ 
& +  \int_\Gamma  C_{jl} \left(\frac{\delta A}{\delta C_{jk}}\nabla_{\Gamma l} \frac{\delta B}{\delta m_{k}} -  \frac{\delta B}{\delta C_{jk} }\nabla_{\Gamma l}\frac{\delta A}{\delta m_{k}} \right)d\vv{x}, \\ 
\end{split}
\end{equation}
where $\delta A/ \delta C_{jk}$ denotes a functional derivative, and $\nabla_\Gamma$ is the surface gradient (which we define more explicitly in Sec.~\ref{sec:Model}). \\
We now define the traceless tensor $\bar{M}$ as $\bar{M}\equiv C -\fr (\mbox{tr}C) P$, where $P$ is the surface projection matrix $P = I - \vv{n} \otimes \vv{n} $. Here $I$ is the bulk unit tensor and $\vv{n}$ is the unit normal on $\Gamma$. For the sake of simplicity we will now assume the interface is flat, which implies that $P$ is a constant tensor, independent of position on the interface. For a curved interface additional terms emerge in the equations, but these are all normal to the interface. Since we are primarily interested in the in-plane stress distribution on the interface, we will not consider these here. Substituting the expression for $M$ in the last equation, we find that the Poisson bracket $\{A,B\}_M$ can be written as
\begin{equation} \label{eq 17}
\begin{split}
\left\{A,B\right\}_M = & \int_\Gamma \left[\bar{M}_{jk}+\fr  (\mbox{tr}C)  \delta_{jk}\right)\nabla_{\Gamma l} \left( [ \delta_{mj}\delta_{nk} - \fr \delta_{jk}\delta_{mn}]   \left\{   \frac{\delta A}{\delta \bar{M}_{mn}} \frac{\delta B}{\delta m_{l}} -  \frac{\delta B}{\delta \bar{M}_{mn} }\frac{\delta A}{\delta m_{l}} \right\} \right)d\vv{x} \\ 
& +  \int_\Gamma \left[\bar{M}_{lk}+\fr  (\mbox{tr}C) \delta_{lk}\right] [ \delta_{mj}\delta_{nk} - \fr \delta_{jk}\delta_{mn}] \left(\frac{\delta A}{\delta \bar{M}_{mn}}\nabla_{\Gamma l} \frac{\delta B}{\delta m_{j}} -  \frac{\delta B}{\delta \bar{M}_{mn} }\nabla_{\Gamma l}\frac{\delta A}{\delta m_{j}} \right)d\vv{x} \\ 
& +  \int_\Gamma \left[ \bar{M}_{jl} +\fr  (\mbox{tr}C) \delta_{jl}\right] [ \delta_{mj}\delta_{nk} - \fr \delta_{jk}\delta_{mn}] \left(\frac{\delta A}{\delta \bar{M}_{mn}}\nabla_{\Gamma l} \frac{\delta B}{\delta m_{k}} -  \frac{\delta B}{\delta \bar{M}_{mn} }\nabla_{\Gamma l}\frac{\delta A}{\delta m_{k}} \right)d\vv{x}.\\ 
\end{split}
\end{equation}
In arriving at this result we used the chain rule to find
\begin{equation}
\label{eq 18}
\frac{\delta A}{\delta C_{jk}} = \frac{\delta A}{\delta \bar{M}_{mn}}   \frac{\delta \bar{M}_{mn}}{\delta C_{jk}}= \frac{\delta A}{\delta \bar{M}_{mn}}    [ \delta_{mj}\delta_{nk} - \fr \delta_{jk}\delta_{mn}].
\end{equation}
Note that strictly speaking the term $\delta_{mj}\delta_{nk}$ should be split as $\fr\delta_{mj}\delta_{nk} + \fr \delta_{mk}\delta_{nj}$, but since both $C$ and $M$ are symmetric tensors this does not affect our final answer. 
If we now focus only on the reversible part of the dynamics associated with $M$, we have
\begin{equation}
\label{eq 19}
\left.\frac{d A}{dt}\right|_{\mbox{\scriptsize rev},M} = \int_\Gamma  \frac{\delta A}{\delta \bar{M}_{jk}}\frac{\partial \bar{M}_{jk}}{\partial t} d\vv{x} = \{A,E\}_M, 
\end{equation}
which we can rewrite as (after integration by parts of the first line of (\ref{eq 17}))
\begin{equation}
\label{eq 20}
 \int_\Gamma  \frac{\delta A}{\delta \bar{M}_{jk}}{\cal H}_{jk} = 0, \\
\end{equation}
with
\begin{equation} \label{eq 21}
{\cal H}_{jk}= \frac{\partial \bar{M}_{jk}}{\partial t} + v_{\Gamma l} \nabla_{\Gamma l} \bar{M}_{jk} - \bar{M}_{lk}\nabla_{\Gamma l}v_{\Gamma j} - \bar{M}_{jl}\nabla_{\Gamma l}v_{\Gamma k} +\delta_{jk}\left( \bar{M}_{il}\nabla_{\Gamma l}v_{\Gamma i} \right) - (\mbox{tr}C)\bar{D}_{\Gamma jk}. 
\end{equation}
Here $D_\Gamma$ is the surface rate of deformation, $D_\Gamma = \frac{1}{2} P \left(\nabla_\Gamma \vv{v}_\Gamma + (\nabla_\Gamma \vv{v}_\Gamma)^T \right) P$ and $\bar{D}_\Gamma$ its traceless component $D-\frac{1}{2} (\tr D) P$. 
Since $A$ was chosen arbitrarily, we must have
\begin{equation} \label{eq 22}
\frac{\partial \bar{M}_{jk}}{\partial t} + v_{\Gamma l} \nabla_{\Gamma l} \bar{M}_{jk} - \bar{M}_{lk}\nabla_{\Gamma l}v_{\Gamma j} - \bar{M}_{jl}\nabla_{\Gamma l}v_{\Gamma k} +\delta_{jk}\left( \bar{M}_{il}\nabla_{\Gamma l}v_{\Gamma i} \right) - (\mbox{tr}C)\bar{D}_{\Gamma jk} =0,
\end{equation}
or in tensor form
\begin{equation} \label{eq 23}
 \partial_{\Gamma , t}^\bullet \bar{M}  -  \nabla_\Gamma\vv{v}_\Gamma \cdot \bar{M} - \bar{M} \cdot (\nabla_\Gamma \vv{v}_\Gamma)^T +P\left(\bar{M}:\nabla_\Gamma\vv{v}_\Gamma\right) - (\mbox{tr}C) \bar{D}_\Gamma =0,
\end{equation}
where $\partial_{\Gamma , t}^\bullet$ denotes the surface material derivative.  Identifying $C=S$ (and hence $\bar{M}=\bar{S}$), we obtain
\begin{equation} \label{eq 24}
 \partial_{\Gamma , t}^\bullet\bar{S}     - \nabla_\Gamma\vv{v}_\Gamma \cdot  \bar{S} -  \bar{S} \cdot (\nabla_\Gamma \vv{v}_\Gamma)^T 
+P\left( \bar{S}:\nabla_\Gamma\vv{v}_\Gamma\right) - (\mbox{tr}S)\bar{D}_\Gamma =0.
\end{equation}
Adding the contributions from the dissipative bracket to this expression, using a simple bi-linear form for the contribution of the tensor to the entropy density \cite{Oettinger2005, Sagis-Oettinger2013}, we obtain
\begin{equation} \label{eq 25}
\partial_{\Gamma , t}^\bullet\bar{S}      - \nabla_\Gamma\vv{v}_\Gamma \cdot  \bar{S} -  \bar{S} \cdot (\nabla_\Gamma \vv{v}_\Gamma)^T 
+P\left( \bar{S}:\nabla_\Gamma\vv{v}_\Gamma\right) - (\mbox{tr}S)\bar{D}_\Gamma + \frac{1}{\tau_S} \bar{S}  = \frac{2\varepsilon_S}{\tau_S}\bar{D}_\Gamma,
\end{equation}
where $\varepsilon_S$ is the surface shear viscosity and $\tau_S$ the surface shear relaxation time. We define the traceless upper-convected surface derivative as
\begin{equation}
    \overline{\overset{\nabla}{\delta}\bar{S}} = \partial_{t, \Gamma}^\bullet \bar{S} - \nabla_\Gamma \vv{v}_\Gamma \bar{S} - \bar{S} (\nabla_\Gamma \vv{v}_\Gamma)^T + P(\bar{S} : \nabla_\Gamma \vv{v}_\Gamma) - \tr (S) \bar{D}_\Gamma. \label{eq:objSurfDeriv}
\end{equation}
So we can rewrite equation \eqref{eq 25} as
\begin{equation} 
\label{eq:devS ODE}
\overline{\overset{\nabla}{\delta}\bar{S}} + \frac{1}{\tau_S} \bar{S}  = \frac{2\varepsilon_S}{\tau_S}\bar{D}_\Gamma.
\end{equation}
which describes the evolution of the traceless part of the surface stress. 

It remains to derive an evolution equation for the trace of the stress. The general bracket for a scalar variable $\Lambda$ is given by \cite[p. 114, Eq. 4.44, last two lines]{Oettinger2005},
\begin{equation}
\begin{split}
\label{eq 26}
\left\{A,B\right\}_\Lambda =& - \int_\Gamma \Lambda\nabla_\Gamma\cdot\left( \frac{\delta A}{\delta \vv{m}} \frac{\delta B}{\delta \Lambda} -  \frac{\delta B}{\delta \vv{m}}\frac{\delta A}{\delta \Lambda} \right)d\vv{x} \\
&+ \int_\Gamma {G}:  \left( \frac{\delta A}{\delta \Lambda}\nabla_\Gamma \frac{\delta B}{\delta \vv{m}} - \frac{\delta B}{\delta \Lambda}\nabla_\Gamma \frac{\delta A}{\delta \vv{m}}  \right) d\vv{x}
\end{split},
\end{equation}
where the second order tensor ${G}$ couples the convective processes for $\Lambda$ to the convective processes of the tensor $S$. Its general form is given by
\begin{equation}
\label{eq 27}
{\bf G} = g_1 S + g_2 P + g_3 S^{-1}.
\end{equation}
The convective dynamics of $\Lambda$ are then given by
\begin{equation}
\label{eq 28}
\left.\frac{d A}{dt}\right|_{\mbox{\scriptsize rev},\Lambda} =  \{A,E\}_\Lambda . 
\end{equation}
Combining this with (\ref{eq 26}), performing an integration by parts of the first term in that expression, and choosing $\Lambda=\mbox{tr}S$, $g_1=2$, and $g_2=g_3=0$, we obtain,
\begin{equation}
\label{eq 30}
\partial_{\Gamma , t}^\bullet \mbox{tr}S  - 2(S:\nabla_\Gamma\vv{v}_\Gamma) =0.
\end{equation}
Adding the dissipative contributions, and switching to $\bar{S}$ for the second term on the left,  we find
\begin{equation}
\label{eq 31}
\partial_{\Gamma , t}^\bullet \mbox{tr}S - 2(\bar{S}:\nabla_\Gamma\vv{v}_\Gamma)  - (\mbox{tr}S)(\mbox{tr}D_\Gamma)
+ \frac{1}{\tau_A}\mbox{tr}S = \frac{2\varepsilon_A}{\tau_A} \mbox{tr}D_\Gamma. 
\end{equation}
The parameters $\varepsilon_A$ and $\tau_A$ are the areal viscosity and the areal relaxation time.

Eqs. \eqref{eq 31} and \eqref{eq:objSurfDeriv}-\eqref{eq:devS ODE} provide the evolution of viscoelastic surface stress tensor under a given flow field. 
Contrary to the previous model from \cite{sagis2010modelling,sagis2011dynamic} the zero trace of $\bar{S}$ is conserved by the equations. 
One can see the equations as a surface-equivalent of the upper convected Maxwell model. 
This analogy becomes apparent if Eq.~\eqref{eq:devS ODE} is added to $\frac{1}{2}P$ multiplied by Eq.~\eqref{eq 31}. 
Evaluating the result for an incompressible surface ($\nabla_\Gamma \cdot\vv{v}_\Gamma=0$) in flat space ($P=$const.) and with $\tau_A=\tau_S$ gives 
\begin{align}
    \partial_{\Gamma , t}^\bullet{S}      - \nabla_\Gamma\vv{v}_\Gamma \cdot  {S} -  {S} \cdot (\nabla_\Gamma \vv{v}_\Gamma)^T 
 +\frac{1}{\tau_S} {S}  = \frac{2\varepsilon_S}{\tau_S}{D}_\Gamma,
\end{align}
which differs from the  usual form of the upper-convected Maxwell equations only by the use of surface operators.

\section{Model of a viscoelastic surface coupled to surrounding fluids}
\label{sec:Model}
\subsection{Governing equations}
\begin{figure}
    \centering
    \begin{subfigure}{0.45\textwidth}
    \centering
        \includegraphics[height=3.5cm]{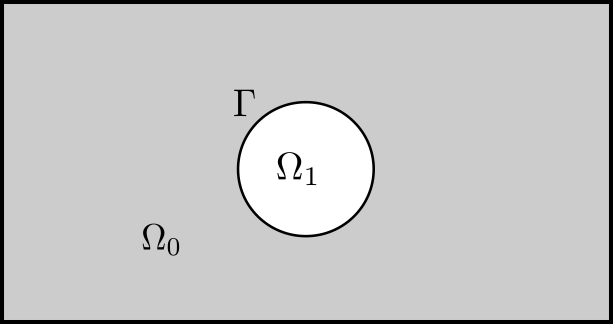}
        \caption{Sketch of the domain.}
        \label{fig:domain2d}
    \end{subfigure}
    \begin{subfigure}{0.5\textwidth}
    \centering
        \includegraphics[height=3.5cm]{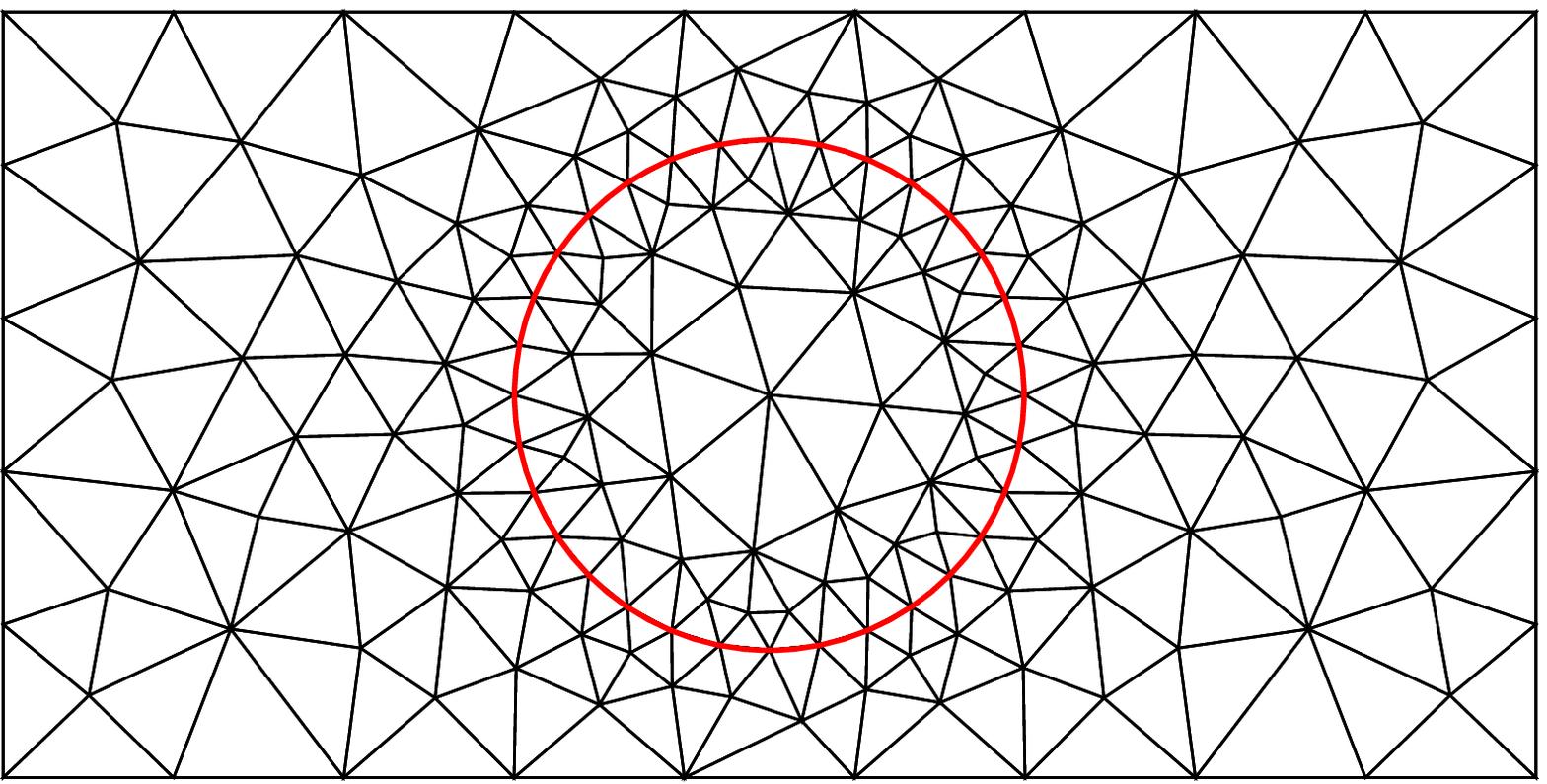}
        \caption{Sketch of the mesh, the surface $\Gamma$ is in red.}
        \label{fig:meshSketch}
    \end{subfigure}
    \caption{Computational setting of a viscoelastic surface embedded in fluids represented in two dimensions.}
    \label{fig1}
\end{figure}

We consider a time dependent viscoelastic surface $\Gamma$ suspended in a fluid with a three dimensional open domain $\Omega$. The domain $\Omega$ consists of two open sub-domains. The internal fluid is labelled $\Omega_1$ and the external fluid $\Omega_0$, such that $\Omega = \Omega_1 \cup \Omega_0$ (see Figure \ref{fig:domain2d}). 

The hydrodynamic system is similar to a  
two-phase flow system with Boussinesq–Scriven surface fluid \cite{bothe2010two}. 
The surrounding fluid on $\Omega$ is modelled by the incompressible Navier-Stokes equation,
\begin{align}
    \nabla \cdot \vv{v} & = 0 & \text{on } \Omega &, \label{eq:stokes1}\\
    \rho \partial_t^\bullet \vv{v} &= -\nabla p + \eta_i \nabla \cdot \left(\nabla \vv{v} + (\nabla \vv{v})^T  \right) & \text{on } \Omega_i & \text{, for } i=0,1, \label{eq:stokes2}
\end{align}
where, $\vv{v}$ is the velocity and $p$ the pressure, $\partial_t^\bullet$ denotes the material derivative. The parameter $\rho$ is the density of the fluid and $\eta_i$ its viscosity on $\Omega_i$. On the surface $\Gamma$ we define the normal $\vv{n}$ and the surface projection matrix $P = I - \vv{n} \otimes \vv{n} $. With this we define the surface gradient and the surface divergence. The surface gradient is $\nabla_\Gamma f = P \nabla f_e$ for a scalar function $f$. $f_e$ is the extension of $f$ outside its domain $\Gamma$. For a vector valued function $\vv{g}$ the surface gradient is $\nabla_\Gamma \vv{g} = \nabla \vv{g}_e P$ and the divergence is $\nabla_\Gamma \cdot \vv{g} = P : \nabla \vv{g}_e$.

Using these definitions the dynamics on the surface are defined. Assuming no slip at the surface, the force balance on $\Gamma$ is \cite{bothe2010two}, 
\begin{equation}
    -\rho_\Gamma \partial_{\Gamma, t}^\bullet \vv{v}_\Gamma + \nabla_\Gamma \cdot (S + \gamma P)= \left[ -p I + \eta_i \left( \nabla \vv{v} + (\nabla \vv{v})^T \right) \right]_0^1 \cdot \vv{n} \qquad \text{on } \Gamma. \label{eq:surfaceForce}
\end{equation}
The first term describes inertial forces on the surface with surface mass density $\rho_\Gamma$. The second term contains the surface stresses, with $S$ being the viscoelastic surface stress obtained from the equations derived in the previous section. 
$\gamma$ is the surface tension.
The term on the right hand side comprises the force which the bulk fluids exerts on the surface. 
The brackets define the interface jump of a quantity by $[f_i]_0^1 = f_1(x_1) - f_0(x_0)$ for a function $f_i \in \Omega_i$, where the points $x_i \in \Omega_i$ are arbitrarily close to $\Gamma$. 
Let us denote that the inertial terms might be negligible if small length scales are considered. For example, most applications of biological cells involve Reynolds numbers $\ll 10^{-2}$. In this case, one may choose $\rho=\rho_\Gamma=0$.
Finally, the change of the surface position is given by, 
\begin{equation}
    \Gamma(t) = \int_0^t \vv{v}_\Gamma(t', \Gamma(t')) dt' + \Gamma_0.
\end{equation}
This equation states that $\Gamma$ is advected with the flow, i.e. we assume no mass flux across the surface. 

\noindent The complete mechanical system of equations is summarized as follows.
\begin{align}
    \nabla \cdot \vv{v} & = 0 & \text{on } \Omega, \label{eq:scaledStokes1} \\
     0 &= -\nabla p + \eta_i \nabla \cdot \left(\nabla \vv{v} + (\nabla \vv{v})^T  \right) & \text{on } \Omega_i &, i=0,1, \label{eq:scaledStokes2} \\
    \bar{S} &= 2 \varepsilon_S \bar{D}_\Gamma - \tau_S \overline{\overset{\nabla}{\delta} \bar{S}}  &\text{on } \Gamma, \label{eq:scaledDevS}\\
    \tr(S) &= 2 \varepsilon_A \tr(D_\Gamma) + \tau_A \left(2(\bar{S}:\nabla_\Gamma \vv{v}_\Gamma) + \tr (S)\tr (D_\Gamma) -\partial_{t, \Gamma}^\bullet \tr S \right) &\text{on } \Gamma, \label{eq:scaledTrS}\\
     -\rho_\Gamma \partial_{t, \Gamma}^\bullet \vv{v}_\Gamma + \nabla_\Gamma \cdot ( S + \gamma P) &= \left[ -p I + \eta_i \left( \nabla \vv{v} + (\nabla \vv{v})^T \right) \right]_0^1 \cdot \vv{n} & \text{on } \Gamma, \label{eq:stressBC} \\
     S &= \bar{S} + \frac{1}{2} (\tr S) P. \label{eq:scaledDevSDef}
\end{align}
Note, that the traceless upper-convected surface derivative used in Eq.~\eqref{eq:scaledDevS} was introduced in Eq.~\eqref{eq:objSurfDeriv}.
The system of equations describes the dynamics of a viscoelastic Maxwell fluid surface immersed in viscous fluids. 

The use of two separate equations for areal ($\tr S$) and shear ($\bar{S}$) stress of the surface allows to distinguish between these two types of deformation and makes it possible to choose large ratios of dilational ($\varepsilon_A, \tau_A$) and shear ($\varepsilon_S, \tau_S$) surface parameters. For example cell cortices have a high resistance to areal deformation, but do allow shear deformation \cite{aland2014diffuse}.

\subsection{Parameters and limit cases}
\label{sec:parameters_and_limits}
The proposed model is a very general description of a viscoelastic in-plane surface fluid and involves several parameters. 
In this section, we will discuss the reduced equations in the limiting cases for special choices of parameters. 

\vspace{0.3cm}
\noindent
{\bf Limit case: Purely viscous surface.}\\
One particularly interesting case is the case of a purely viscous fluid surface. In this case the relaxation of elastic stresses is so fast (i.e. $\tau_S,\tau_A \rightarrow 0$) that no elastic stress accumulates. Consequently, the three equations \eqref{eq:scaledDevS},\eqref{eq:scaledTrS} and \eqref{eq:scaledDevSDef} are reduced to $S=2\varepsilon_S \bar{D}_\Gamma + \varepsilon_A {\rm tr}(D_\Gamma)P$. 
In the absence of bulk fluids the system \eqref{eq:scaledStokes1}-\eqref{eq:scaledDevSDef} reduces to a single equation for the surface momentum balance:
\begin{align}
-\rho_\Gamma \partial_{t, \Gamma}^\bullet \vv{v}_\Gamma + \nabla_\Gamma \cdot (2\varepsilon_S \bar{D}_\Gamma + (\varepsilon_A {\rm tr}(D_\Gamma) + \gamma) P) &= 0. \label{eq: purely viscous surface limit}
\end{align}

\vspace{0.3cm}
\noindent
{\bf Limit case: Viscous incompressible surface.}\\
In the last decade several first numerical models were proposed to simulate fluidic incompressible surfaces \cite{reuther2018solving, reuther2016incompressible, jankuhn2019higher, reusken2020stream, ArroyoDeSimone_PRE_2009}. 
The requirement of areal incompressibility, $\nabla_\Gamma \cdot \vv{v}_\Gamma = 0$, makes the surface dilational part of Eq.~\eqref{eq: purely viscous surface limit} obsolete, as $\tr D_\Gamma = 0$. The remaining dilational part only involves the surface tension $\gamma$ which now assumes the role of a Lagrange multiplier providing the surface dilational force to keep the interface area locally conserved. Hence, $\gamma$ is not a free parameter anymore. It assumes a similar role as the pressure in a bulk Navier-Stokes equation, and can be interpreted as a surface pressure. Further using $\nabla_\Gamma \cdot \vv{v}_\Gamma = 0$, the surface rate of deformation is trace-free, i.e. $\bar{D}_\Gamma=D_\Gamma$. With these substitutions in the model the complete surface Navier-Stokes system becomes
\begin{align}
-\rho_\Gamma \partial_{t, \Gamma}^\bullet \vv{v}_\Gamma + \nabla_\Gamma \cdot (2\varepsilon_S {D}_\Gamma + \gamma P) &= 0, \label{eq: blabla} \\ 
\nabla_\Gamma\cdot \vv{v}_\Gamma &= 0. 
\end{align}
Hence, we recover the model of a viscous incompressible surface \cite{jankuhn2018incompressible}. The only two remaining parameters are the surface density $\rho_\Gamma$ and surface shear viscosity $\varepsilon_S$.

\vspace{0.3cm}
\noindent
{\bf Limit case: Purely elastic surface.}\\
When both the viscosity and relaxation time are very large, viscous effects become less and less important, such that the interface rheology approaches a purely elastic material.
Dividing Eq.~\eqref{eq:scaledDevS} by $\tau_S$ and Eq.~\eqref{eq:scaledTrS} by $\tau_A$ and letting both relaxation times tend to infinity, we obtain
\begin{align}
0 &= 2 K_S \bar{D}_\Gamma -  \overline{\overset{\nabla}{\delta} \bar{S}}  &\text{on } \Gamma, \label{eq:scaledDevS2}\\
0 &= 2 K_A \tr(D_\Gamma) + 2(\bar{S}:\nabla_\Gamma \vv{v}_\Gamma) + \tr (S)\tr (D_\Gamma) -\partial_{t, \Gamma}^\bullet \tr S &\text{on } \Gamma, \label{eq:scaledTrS 2}
\end{align}
where we have introduced the surface area dilation modulus $K_A=\varepsilon_A / \tau_A$ and the surface shear modulus $K_S=\varepsilon_S /\tau_S$. 
Adding Eq.~\eqref{eq:scaledDevS2} to $\frac{1}{2}P$ times Eq.~\eqref{eq:scaledTrS 2} gives a single evolution equation for the surface stress. 
The equation can be simplified by use of $\partial_{t, \Gamma}^\bullet P 
= \nabla_\Gamma\vv{v} + (\nabla_\Gamma \vv{v})^T - 2D_\Gamma$, which can be derived from $\partial_{t, \Gamma}^\bullet {\bf n}
= -{\bf n}\nabla_\Gamma\vv{v}$ (Lemma 37 in \cite{barrett2020parametric}), so that we arrive at
\begin{align}
\partial_{t, \Gamma}^\bullet S -\nabla_\Gamma \vv{v}S-S\nabla_\Gamma \vv{v}^T &= 2 K_S \bar{D}_\Gamma + K_A \tr(D_\Gamma) P.\label{eq: common elastic stress}
\end{align}
Equation~\eqref{eq: common elastic stress} defines an evolution equation for the in-plane stress of an elastic surface. The left-hand side is the surface upper-convected derivative of $S$. 

\vspace{0.3cm}
\noindent
{\bf Including bending stiffness.}\\
Our surface rheological model describes the in-plane contribution of viscoelastic surfaces. The out-of-plane contribution, i.e. the bending stiffness, has been omitted so far. 
In fact, for very thin closed surfaces (compared to the size of the considered object) bending stiffness becomes negligible as it scales cubically with the interface thickness (see e.g. Landau and Lifshitz \cite{land86}).
However, in certain scenarios bending contributions become important, for example, to describe biomembranes and red blood cells, e.g. \cite{ong2020immersed,veerapaneni2009boundary,aland2014diffuse,Marth2015,lowengrub2016numerical}. 
In this case the bending stiffness force can be easily added to the interfacial stress balance by modifying Eq.~\eqref{eq:stressBC} to 
\begin{align}
-\rho_\Gamma \partial_{t, \Gamma}^\bullet \vv{v}_\Gamma + \nabla_\Gamma \cdot ( S + \gamma P) =& \left[ -p I + \eta_i \left( \nabla \vv{v} + (\nabla \vv{v})^T \right) \right]_0^1 \cdot \vv{n} \label{eq: with bending} \\
&+ 2K_B \left[\Delta_\Gamma(H-H_0)+(H-H_0)\|\nabla_\Gamma {\bf n}\|^2 - \frac{1}{2}H(H-H_0)^2 \right]{\bf n} & \text{on } \Gamma,
\end{align}
where $K_B$ is the bending stiffness, $H$ is the total curvature and $H_0$ is the spontaneous curvature of the surface. We refer to \cite{lowengrub2016numerical} for a derivation of the used bending stiffness force.

\vspace{0.3cm}
\noindent
{\bf Choice of parameters.}\\
The surface rheology in the full model is governed by six parameters: $\rho_\Gamma, \varepsilon_A, \varepsilon_S, \tau_A, \tau_S, \gamma$. 
The surface density $\rho_\Gamma$ is typically easy to compute as the product of the 3D density of the interface material and thickness $h$. 
The other parameters can in principle be measured, for example by oscillating bubble methods or bubble pressure tensiometry. In these methods bubbles or droplets are deformed and their shape response or pressure is measured. We refer to the review paper \cite{sagis2011dynamic}, for a comprehensive overview on interface rheology of various systems. 
For interfaces at the microscale, e.g. the cell cortex, measurements are more tricky and often involve matching of experimental data with numerical simulation results, see e.g. \cite{mokbel2020poisson}.  

In case of a purely elastic interface the parameters can be reformulated to surface elastic parameters. These can be computed from elastic parameters of the 3D interface material, i.e. from Young's modulus $E$, Poisson's ratio $\nu$ and surface thickness $h$: 
\begin{align*}
K_{A}=\frac{h E}{2(1-\nu)}, \quad K_{S}=\frac{h E}{2(1+\nu)}, \quad K_{B}=\frac{h^{3} E}{24\left(1-\nu^{2}\right)}.
\end{align*}
These parameters describe the surface area dilation modulus ($K_A$), surface shear modulus ($K_S$) and bending stiffness ($K_B$) and have been used in Eqs.~\eqref{eq:scaledDevS2},\eqref{eq:scaledTrS 2} and \eqref{eq: with bending}.

\section{Discretization}
\label{sec:Discretization}
We end up with a coupled bulk/surface system on time-dependent domains. 
The stable discretization of such a system is a challenging task for which we develop a Finite Element scheme in the following. The scheme is based on connected numerical grids for the bulk domains. The boundaries of the inner fluid $\Omega_1$ and the inner boundaries of the outer fluid $\Omega_0$ have the same grid points as the surface grid, see Fig.~\ref{fig:meshSketch}.

\subsection{ALE-approach}
\label{sec:AleApproach}
For the implementation of the surface in the fluid we use an ALE-approach (Arbitrary Lagarangian-Eulerian approach). Thereby, we introduce the surface grid velocity $\vv{w}_\Gamma$. While the surface $\Gamma$ needs to move with the flow in normal direction, tangential motion can be chosen arbitrarily. 
We propose
\begin{equation}
    \vv{w}_\Gamma = (\vv{v}_\Gamma \cdot \vv{n}) \vv{n} + P \vv{v}_{\text{avg}} \label{eq:vgrid},
\end{equation}
where $\vv{v}_{\text{avg}}$ is the average velocity of the surface and is defined as $\vv{v}_{\text{avg}} = \frac{1}{|\Omega_1|} \int_{\Omega_1} \vv{v} d\Omega_1 $. 
This form not only moves the surface in the correct way, $\vv{w}_\Gamma\cdot{\bf n} = \vv{v}_\Gamma\cdot{\bf n}$, but also ensures a nice distribution of grid points along the surface. For example, for a purely tangential flow it ensures $\vv{w}_\Gamma=0$, and for a purely translational flow ($\vv{v}= \vv{v}_{avg}$), it ensures $\vv{w}=\vv{v}_{avg}$, such that in these cases the grid spacing is perfectly conserved. 

The grid velocity in the fluid domains, $\vv{w}$ is calculated by harmonically extending the surface velocity $\vv{w}_\Gamma$ to the domain $\Omega$ by solving the following Laplace equation,
\begin{align}
\begin{split}
\Delta \vv{w} &= 0 \qquad \text{on } \Omega, \\
\vv{w} &= \vv{w}_\Gamma \qquad \text{on } \Gamma, \\
\vv{w} &= \vv{0} \qquad \text{on } \partial \Omega \setminus \Gamma.
\label{eq:Laplace}
\end{split}
\end{align}
The surface grid velocity is subtracted from the velocity in the definition of the material derivative, hence for a function $f$ on $\Gamma$ it becomes, 
\[
\partial_t^\bullet f = \pdv{f}{t} + ((\vv{v}_\Gamma - \vv{w}_\Gamma) \cdot \nabla_\Gamma ) f.
\]
where $\pdv{f}{t}$ is the time derivative along a moving grid point.
A similar definition holds for the material derivative in $\Omega$.

\subsection{Time integration}
\label{sec:timeIntegration}
The time integration is done with adaptive time step sizes, dependent on the residual error and the change in $\tr S$, $\bar{S}$ and $\vv{v}$.
The complete bulk/surface system on the moving domains is decoupled into three subsystems which are solved subsequently during each time step.
Each time step from time $t^n$ to $t^{n+1}$ the equations \eqref{eq:scaledStokes1} - \eqref{eq:scaledDevSDef} are solved in the following order.
\begin{enumerate}
    \item The new surface stress $S^{n+1}$ is calculated by solving equations \eqref{eq:scaledDevS} and \eqref{eq:scaledTrS}. For this we use the current surface $\Gamma^n$ and the velocities $\vv{v}^n$ and $\vv{w}_\Gamma^n$ from the previous time step. The weak form of equations \eqref{eq:scaledTrS} and \eqref{eq:scaledDevS} is given below in \eqref{eq:trSweak} and \eqref{eq:devSweak}.
    An IMEX scheme with Euler time integration is used, the l.h.s. of equation \eqref{eq:trSweak} and \eqref{eq:devSweak} is solved implicitly, the remaining parts explicitly. 

    \item The Navier-Stokes equations (\eqref{eq:scaledStokes1}, \eqref{eq:scaledStokes2}) are solved on the fixed  grid $\Omega^n$ from the last time step. We use a semi-implicit Euler method, where only the nonlinear term is treated semi-implicitly. The stress boundary condition \eqref{eq:stressBC}  is incorporated using the previously computed surface stress $S^{n+1}$ on $\Gamma^n$. Their weak form is given in equation \eqref{eq:stokesWeak}.
    
    \item Finally, the grids are advected. Thereby, the surface grid velocity $\vv{w}_\Gamma^{n+1}$ is computed by  Eq.~\eqref{eq:vgrid} using $\vv{v}^{n+1}$. 
    The bulk grid velocity $\vv{w}^{n+1}$ is computed as its harmonic extension by solving the system of equations \eqref{eq:Laplace}.
    Grid points of the surface and bulk grid are then displaced by $(t^{n+1}-t^n)\vv{w}_\Gamma^{n+1}$ and $(t^{n+1}-t^n)\vv{w}^{n+1}$, respectively, to obtain $\Gamma^{n+1}$ and $\Omega^{n+1}$.  
    
\end{enumerate}

\subsection{Spatial discretization}
\label{sec:spatial discretization}
For the spatial discretization we use the Finite Element method, which is implemented using the C++ library AMDiS \cite{VeyVoigt_CVS_2007, Witkowski2015}. The grid is separated into three meshes. Two three dimensional meshes for the Stokes equation and one curved two dimensional mesh for the surface stress. The discretization on the fluid domains $\Omega_0$ and $\Omega_1$ will be referred to as $T_0$ and $T_1$ respectively. The mesh on the surface $\Gamma$ is named $T_\Gamma$. The meshes are constructed such that they match at the interface. That is, each gridpoint in $T_\Gamma$ corresponds to a gridpoint on $T_1$ located on $\partial \Omega_1$ and a gridpoint in $T_0$ located on the internal boundary of $\Omega_0$ (see Figure \ref{fig:meshSketch}). Because of the corresponding gridpoints of $T_0$ and $T_1$ we can implement the jump in pressure  and the continuous velocity across the boundary (for a more detailed explanation see the Appendix in \cite{mokbel2020ale}).\\ 
\\
For the finite element spaces of the fluid we use first order polynomials for the pressure and second order polynomials for the velocity. So the finite element spaces for the pressure and the velocity are,
\begin{align}
    P_{i} &= \Big\{ q \in C^0(\bar{\Omega}_i) \cap L^2(\Omega_i) \Big| \evalat[\big]{q}{k} \in P_1(k), k \in T_i \Big\} \text{ for } i=0,1, \\
    V &= \Big\{ u \in C^0(\bar{\Omega}) \cap H^1(\Omega) \Big| \evalat[\big]{v}{k} \in P_2(k), k \in T_0 \cup T_1 \Big\}.
\end{align}
This is an extension of the Taylor-Hood finite element space, as it results in more degrees of freedom for the pressure on the mesh at the surface. We chose for Taylor-Hood because it gives the optimal convergence rate for these low order elements \cite{LongChen}. The weak form of the Stokes equations then reads: Find $(\vv{v}^{n+1}, p_0^{n+1}, p_1^{n+1}) V \times \in P_{0} \times P_{1}$ such that
\begin{equation}
    \begin{aligned}
        0 &= \int_\Omega \left( \rho \frac{\vv{v}^{n+1} - \vv{v}^n}{t^{n+1}-t^n} - \left((\vv{v}^n-\vv{w}^n) \cdot \nabla\right) \vv{v}^{n+1}\right) \cdot \vv{u} ~d \vv{x} \\
        &+ \int_{\Omega_0} \left( \eta_0 (\nabla \vv{v}^{n+1} + (\nabla \vv{v}^{n+1})^T):\nabla \vv{u} - p_0^{n+1} \nabla \cdot \vv{u} + q_0 \nabla \cdot \vv{v}^{n+1} \right) d\vv{x} \\
        &+ \int_{\Omega_1} \left( \eta_1 (\nabla \vv{v}^{n+1} + (\nabla \vv{v}^{n+1})^T):\nabla \vv{u} - p_1^{n+1} \nabla \cdot \vv{u} + q_1 \nabla \cdot \vv{v}^{n+1} \right) d\vv{x} \\
        &+ \int_{\Gamma^n} \left( \rho_\Gamma \left( \frac{\vv{v}_\Gamma^{n+1} - \vv{v}_\Gamma^n}{t^{n+1}-t^n} - (\vv{v}_\Gamma^n-\vv{w}_\Gamma^n) \cdot \nabla_\Gamma \vv{v}_\Gamma^{n+1}\right) + \nabla_\Gamma \cdot (S^{n+1} + \gamma P^n) \right) \cdot \vv{u} ~d\vv{x}
    \end{aligned}
    \label{eq:stokesWeak}
\end{equation}
holds for all $(\vv{u}, q_0, q_1) \in V \times P_{0} \times P_{1}$. The superscript $n$ denotes the time step. The finite element space of the surface is first order as well, such that this space is of the same order as the space for the pressure. So it reads, 
\begin{equation}
    P_\Gamma = \Big\{ \phi \in C(\Gamma) \cap L^2(\Gamma) \Big| \evalat[\big]{\phi}{k} \in P_1(k), k \in T_\Gamma \Big\}.
\end{equation}
We found that the convective terms in equations \eqref{eq:scaledTrS} and \eqref{eq:scaledDevS} resulted in small numerical oscillations. To dampen these we added a numerical diffusion term. 
The weak forms of the surface stress equations then become: Find $\tr S^{n+1} \in P_\Gamma$ and $\bar{S}_{ij} \in P_\Gamma$ such that the following equations hold, 
\begin{equation}
\begin{aligned}
    &\int_{\Gamma^n} \left(\tr S^{n+1}\left(1 + \frac{\tau_A}{t^{n+1}-t^n}\right) \phi + D_{num} \nabla_\Gamma (\tr S^{n+1}) \cdot \nabla_\Gamma \phi\right) d \vv{x} \\
    &= \int_{\Gamma^n} \left( 2 \varepsilon_A \tr D_\Gamma^n + \tau_A\left( 2(\bar{S}^n:\nabla_\Gamma \vv{v}^n_\Gamma) + \tr (S^n)\tr (D^n_\Gamma) - \left((\vv{v}_\Gamma^n-\vv{w}_\Gamma^n)\cdot \nabla_\Gamma  \right) \tr S^{n} + \frac{\tr S^n}{\Delta t} \right) \right) \phi ~d\vv{x},
\end{aligned}
\label{eq:trSweak}
\end{equation}
\begin{equation}
\begin{aligned}
    &\int_{\Gamma^n} \left(\bar{S}_{ij}^{n+1}\left(1 + \frac{\tau_S}{t^{n+1}-t^n}\right) \phi + D_{num} \nabla_\Gamma (\bar{S}_{ij}^{n+1}) \cdot \nabla_\Gamma \phi\right) d \vv{x} \\
    &= \int_{\Gamma^n} \left( P^n \left( 2 \varepsilon_S \bar{D}_{\Gamma}^n - \tau_S\left(\left((\vv{v}_\Gamma^n-\vv{w}_\Gamma^n)\cdot \nabla_\Gamma \right) \bar{S}^n -  \frac{\bar{S}^n}{\Delta t} - \nabla_\Gamma \vv{v}_\Gamma^n \bar{S}^n - (\nabla_\Gamma \vv{v}_\Gamma^n)^T \bar{S}^n \right) \right) P^n \right)_{ij}\phi d\vv{x}\\
    &- \int_{\Gamma^n} \tau_S \left( P^n(\bar{S}:\nabla_\Gamma \vv{v}_\Gamma^n) - \tr(S^n) \bar{D}^n_{\Gamma}  \right)_{ij}\phi d\vv{x},
\end{aligned}
\label{eq:devSweak}
\end{equation}
for all $\phi \in P_\Gamma$ and $i,j \in \{1,2,3\}$. Where $D_{num}$ is the numerical diffusion coefficient. For the simulations in Sections \ref{sec:surfaceInShearFlow} and \ref{sec:outlook} we chose $D_{num} = 0.01$, for the remaining simulations $D_{num} = 0$. 

Note, that we have contracted the r.h.s. of Eq.~\eqref{eq:devSweak} with the projection matrix, i.e., it is multiplied by $P^n$ from both sides. This is consistent as the corresponding terms are tangential by definition. But in the discrete scheme the terms are only tangential for the grid on which they were computed (time step $n-1$). So the contraction with $P$ assures that such errors do not accumulate and that $\bar{S}$ remains tangential to $\Gamma$.

By the definition of the traceless surface stress $\bar{S}$ we know that its trace should be 0. So any nonzero value for $\tr \bar{S}$ can be seen as a numerical error. Therefore we also make $\bar{S}$ traceless at the end of each iteration by $\bar{S} := \bar{S} - \frac{\tr \bar{S}}{2} P$.

\section{Numerical results}
\label{sec:num}

To validate our model and to demonstrate our motivation for choosing these equations to express the surface stress, we simulate two toy problems which admit analytical solutions. First, we validate the areal surface stress \eqref{eq:scaledTrS} by considering the surface stress for periodic inflation and deflation of a spherical surface. Secondly, we demonstrate the connection of the model with a two dimensional Maxwell material which can be used to validate the numerical solution of the shear surface stress equation \eqref{eq:scaledDevS}.

\subsection{Validation 1: Stretching of a sphere}
To validate the equation for the areal stresses \eqref{eq:scaledTrS} and to demonstrate the influence of including both viscosity and elasticity in the model, a toy problem will be solved analytically. For this we only consider the surface and ignore the fluid in $\Omega$. So only equations \eqref{eq:scaledDevS} and \eqref{eq:scaledTrS} have to be solved. The initial shape of the surface is the unit sphere centred at the origin and we set the velocity field as $\vv{v} = c_1 \sin(c_2 t)\vv{x}$, where $c_1$, $c_2$ are constants in $\mathbb{R}$.  This velocity field corresponds to a periodic inflation/deflation of the sphere. The initial value of $S$ is set to $0$, so the shape of the surface at $t=0$ is its undeformed state. For this velocity field the surface only dilates and there is no shear deformation, hence $\bar{S}=0$. Therefore we only need to solve equation \eqref{eq:scaledTrS}. 

\begin{figure}
    \centering
    \includegraphics[width=0.9\textwidth]{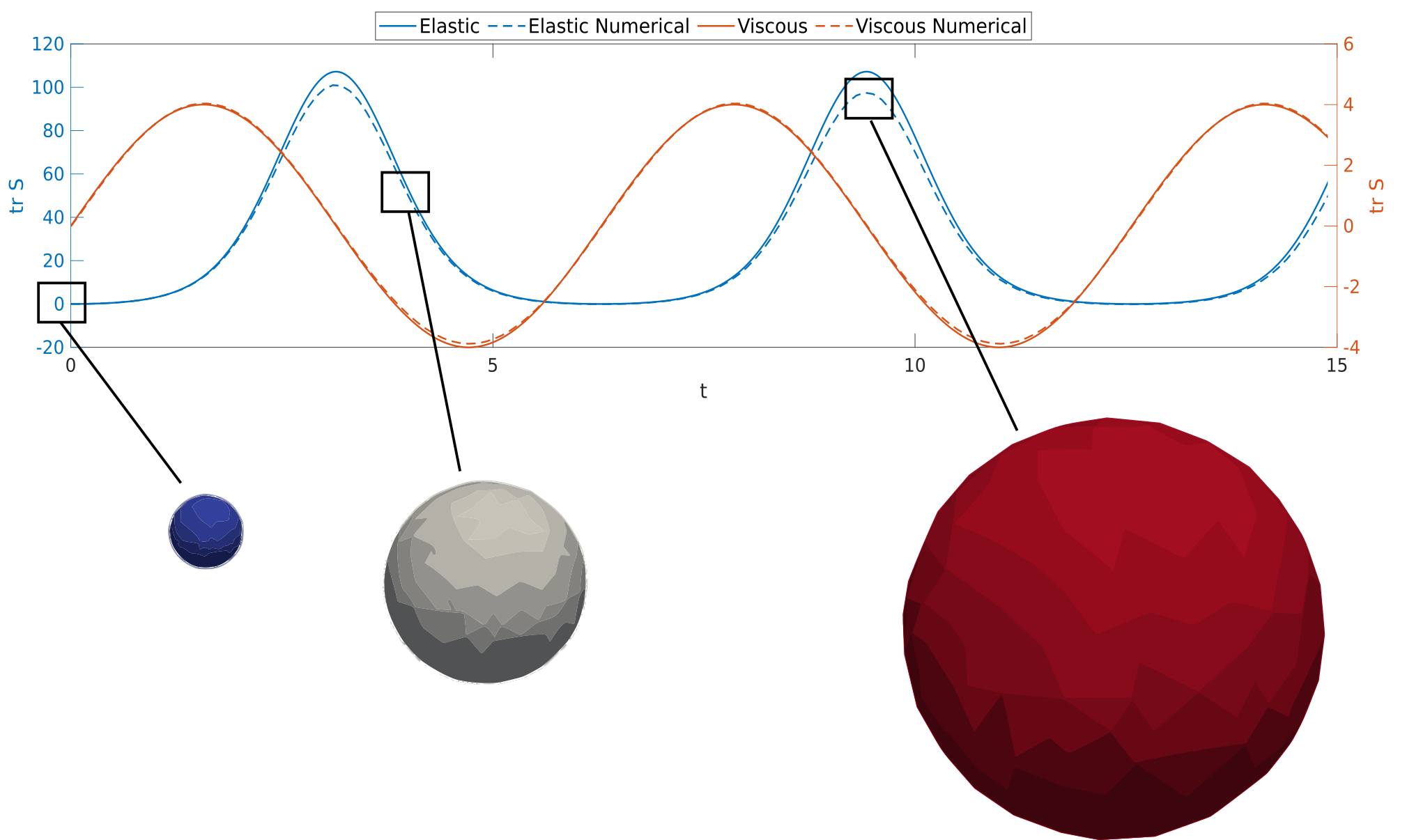}
    \caption{The trace of $S$ for the purely viscous and purely elastic cases. Analytical solutions correspond to Eqs.~\eqref{eq:solution_viscous} and \eqref{eq:solution_elastic} with  $c_1= c_2=1$. Numerical simulations were conducted with $\varepsilon_A = \tau_A = 10^3$ for the elastic case and $\varepsilon_A = 1$ and $\tau_A = 0.01$ for the viscous case. The spheres below describe the size of the sphere while it is being inflated and deflated. The colour represents the value of the surface stress for the simulation of the elastic case. Red for a high surface stress, blue for low.}
    \label{fig:stretchingSphere}
\end{figure}

To compare the numerical solution of Eq.~\eqref{eq:scaledTrS} we develop an analytical solution. For the given velocity field, the grid velocity $\vv{w}_\Gamma$, as defined in equation \eqref{eq:vgrid} is equal to the velocity of the surface $\vv{v}$. This reduces the material derivative $\partial_t^\bullet \tr S$ to a regular time derivative $\partial_t \tr S$ along a moving grid point, reducing equation \eqref{eq:scaledTrS} to,
\begin{equation}
\tr(S) = 4 \varepsilon_A  c_1 \sin(c_2 t) + \tau_A \left( 2 c_1 \tr (S) \sin(c_2 t) -\partial_t \tr S \right).
\label{eq:ODEstretching}
\end{equation}
We distinguish two asymptotic cases, the purely elastic and the purely viscous case.
\begin{enumerate}

    \item For the elastic case the parameters are $\tau_A, \varepsilon_A \gg 1$. For the choice $\tau_A=\varepsilon_A$, equation \eqref{eq:ODEstretching} reduces to the ODE,
    \begin{equation}
        0 = 4 c_1 \sin(c_2 t) + 2 c_1 \text{tr}(S) \sin(c_2 t) - \partial_t S.
    \end{equation}
    Which has the following solution,
    \begin{equation}
        \tr S = 2 \left( \exp \left( \frac{2 c_1}{c_2}(1-\cos(c_2 t)) \right) - 1\right). \label{eq:solution_elastic}
    \end{equation}
    Hence the surface stress is a periodic function and in anti-phase with the surface rate of deformation and equivalently in phase with the surface deformation, which can be found by integrating the rate of deformation w.r.t. time $t$.
    
    \item For the viscous case it holds that $\tau_A \ll 1$ and $\varepsilon_A \sim 1$. So for this case the solution can be approximated by 
    \begin{align} 
       \tr S = 4 c_1 \varepsilon_A \sin (c_2 t). \label{eq:solution_viscous}
    \end{align}
    Hence the surface stress is in phase with the surface rate of deformation. 
    
\end{enumerate}
Both analytical solutions, \eqref{eq:solution_elastic} and \eqref{eq:solution_viscous}, are plotted in Figure \ref{fig:stretchingSphere}, including the numerical solutions found with our model. For the viscous case, we find nearly perfect agreement between numerical and analytical results. Also, for the elastic case, the numerical solution is close to the analytic ones, but the difference increases over time. This is caused by the viscous dissipation of the stress.
Analytical and numerical solutions illustrate the well known phase behavior between deformation and stress being in phase (elastic) or in anti-phase (viscous), and with viscoelastic behavior characterized by being between these extremes \cite{Phan-Thien2013}. 

\subsection{Validation 2: Convergence towards a Maxwell fluid surface}
\label{subsec:UCD}
\begin{figure}
    \centering
    \begin{subfigure}{0.45\textwidth}
    \centering
        \includegraphics[width=0.95\textwidth]{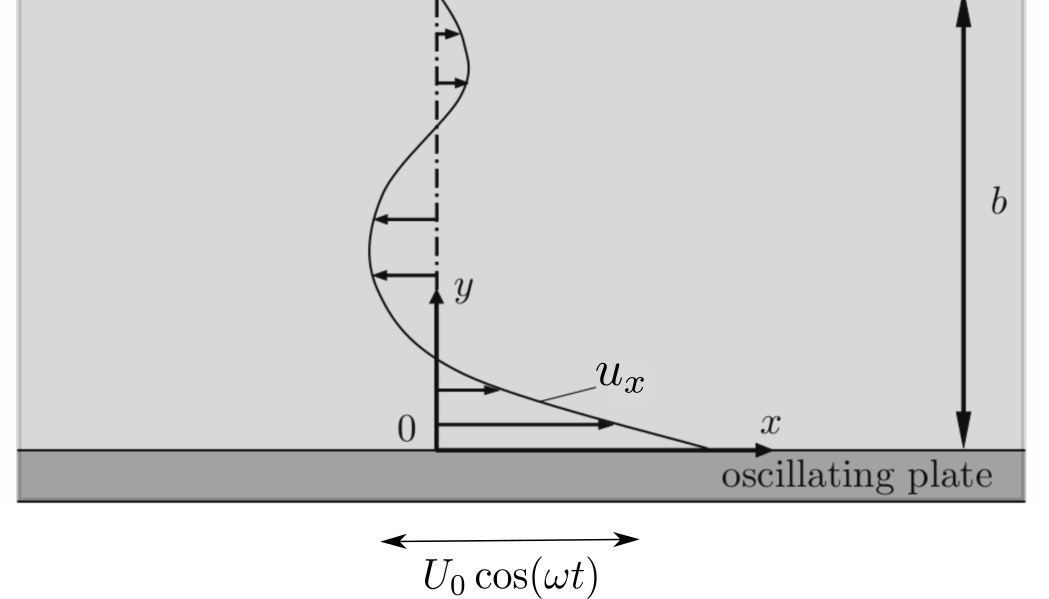}
        \caption{Domain of the two dimensional Maxwell material, derived from  \cite{BoleMa}.}
        \label{fig:BoleMa}
    \end{subfigure}
    \begin{subfigure}{0.45\textwidth}
    \centering
        \includegraphics[width=0.95\textwidth]{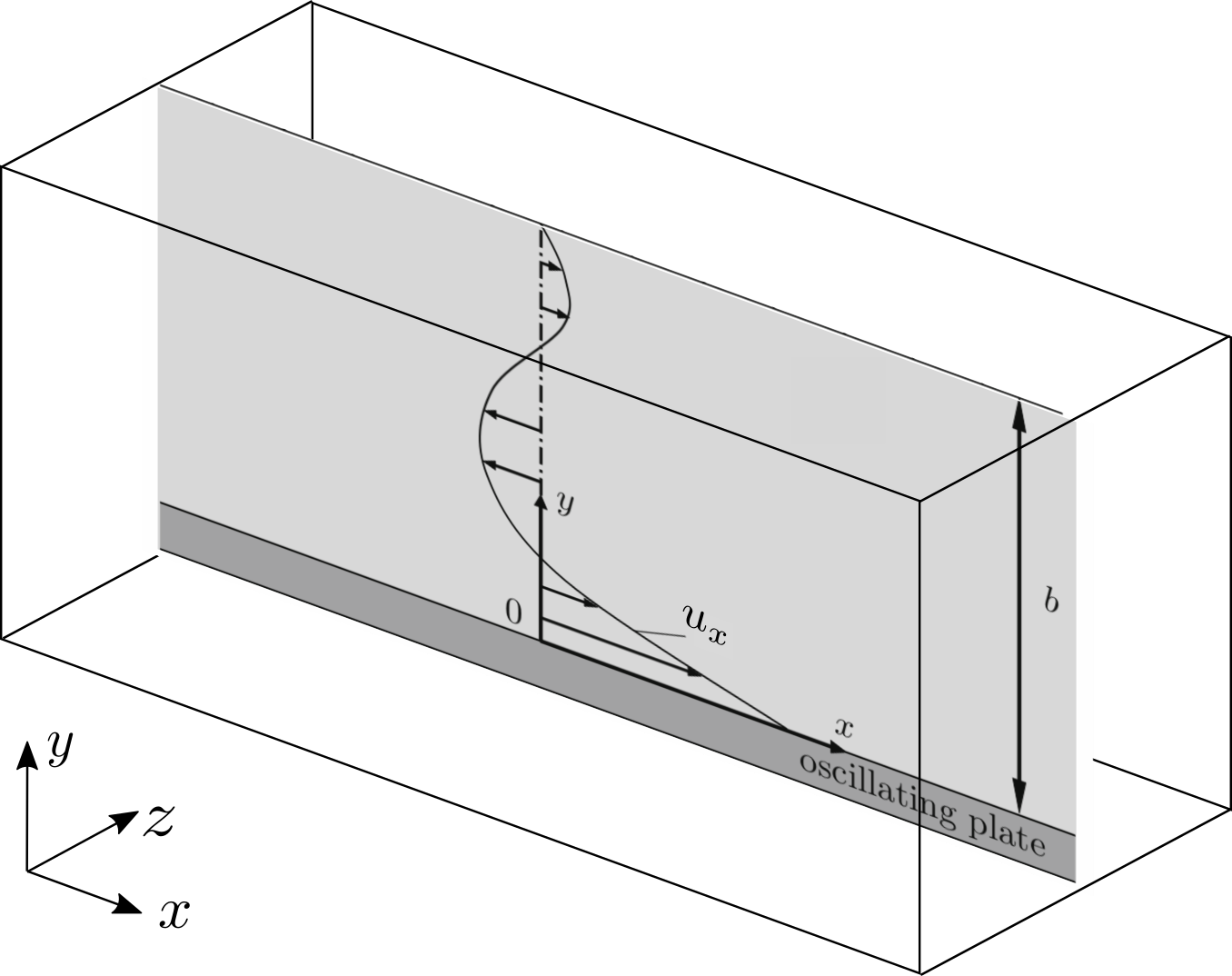}
        \caption{Suspension of the two dimensional surface (grey) in a three dimensional fluid.}
        \label{fig:BoleMa3D}
    \end{subfigure}
    \caption{Graphical representation of the domains.}
\end{figure}

To compare our surface stress model to a Maxwell material we use a two dimensional flat surface $\Gamma$ that is suspended in a three dimensional fluid $\Omega$. The numerical solutions will then be compared to the analytic solution of a two dimensional incompressible Maxwell fluid given by B. Ma et al. in \cite{BoleMa}. They consider a Maxwell material 
between two parallel plates at $y=0$ and $y=b$, as is shown in Figure \ref{fig:BoleMa}. The upper plate remains stationary, the lower moves periodically with period $\omega$ in the $x$-direction, which drives the flow. The governing equations for the two-dimensional Maxwell fluid in the $x$, $y$-plane are \cite{BoleMa},
\begin{equation}
\begin{aligned}
    \nabla \cdot \vv{v} &= 0, \\
    \rho_\Gamma \partial_t^\bullet \vv{v}_\Gamma &= -\nabla p + \nabla \cdot S, \\
    S &= 2\varepsilon_S D - \tau_S \overset{\nabla}{\delta} S, \\
    v_x &= U_0 \cos (\omega t) \qquad \text{at } x=0.
\end{aligned}
\label{eq:BoleMaUnscaled}
\end{equation}
where we have adapted the names of the variables to coincide with our notation. 
The two-dimensional fluid is incompressible, so there is only shear and no dilational stress.

To test our numerical method, we embed the above, two-dimensional Maxwell fluid, into three dimensional space by augmenting the $z$-direction. Hence, the Maxwell fluid is imposed as a surface $\Gamma$ in the middle (at $z=0$) of the three dimensional computational domain $\Omega$, see Fig.~\ref{fig:BoleMa3D}. 
Assuming the 3D-space filled with viscous fluids, we obtain the setting of our proposed numerical model. The boundaries are set at a distance of 1 from the surface.
By opposing no conditions on both boundaries parallel to the surface, we obtain a mirror symmetry in the system at $z=0$. This implies that the surface must remain stationary at $z=0$ and that the pressure jump across the surface is zero. 
As no flow in the $z$-direction should emerge, $\nabla \cdot \vv{v} = \nabla_\Gamma \cdot \vv{v} = 0$. 
Hence the flow dynamics at the surface should approach those of the two-dimensional Maxwell fluid, given the viscosity of the surrounding fluids is low. 

\begin{figure}[h]
    \centering
    \begin{subfigure}{\textwidth}
    \centering
        \includegraphics[width = \textwidth]{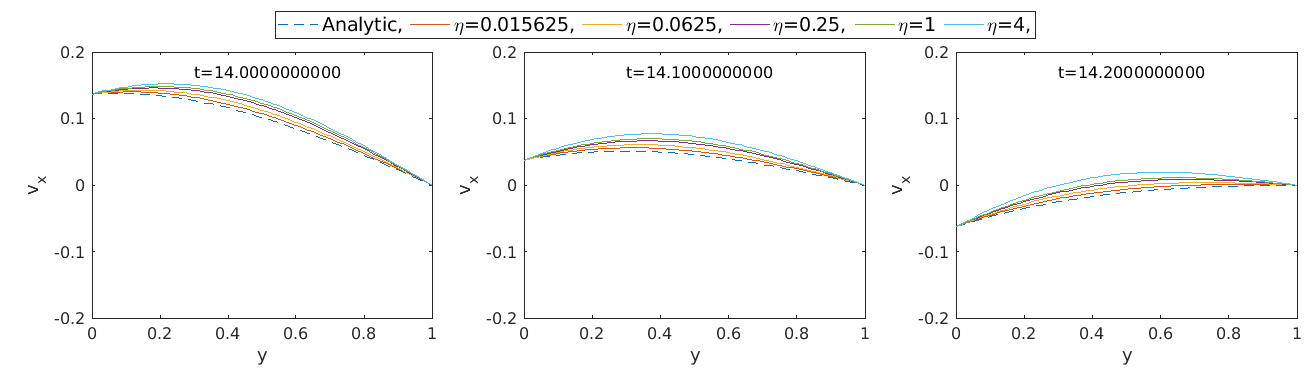}
        \caption{$\varepsilon_S=10$}
        \label{fig:MScompareToAna1}
    \end{subfigure}
    \begin{subfigure}{\textwidth}
    \centering
        \includegraphics[width = \textwidth]{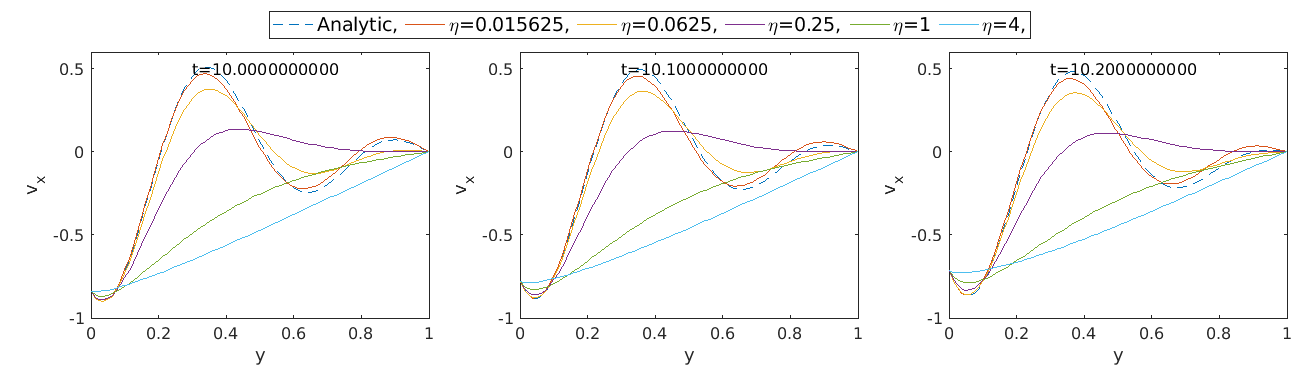}
        \caption{$\varepsilon_S=0.1$}
        \label{fig:MScompareToAna2}
    \end{subfigure}
    \caption{Two examples of the numerical solutions for the surface suspended in a fluid, compared to the analytic solutions of a two dimensional Maxwell material at three different times. $\omega = \pi$ and the remaining parameters are 1.}
    \label{fig:MScompareToAna}
\end{figure}

We are therefore in a position to compare the analytical solution of a two-dimensional Maxwell fluid proposed in \cite{BoleMa} with the numerical solution of our three-dimensional surface/bulk model. 
To this end, the surface tension is set to $\gamma=0$ and we chose $\tau_S$ and $\varepsilon_S$ such that we are well in the viscoelastic regime. The choice of the dilational parameters is arbitrary as no surface dilation should occur. Here we use the same values, $\varepsilon_A=\varepsilon_S$ and $\tau_A=\tau_S$. Because of the symmetry w.r.t. the $x$-axis we only have to calculate the solution for one arbitrary value of $x$. 
In our simulations we observe that decreasing the fluid viscosity $\eta$, reduces the differences between the analytic and numerical solutions. This can be seen in the two examples in Figure \ref{fig:MScompareToAna}. For these simulations we chose $\varepsilon_S=0.1$ or $\varepsilon_S = 10$, $\tau_S=1$, $\omega=\pi$ and the remaining parameters equal to 1. By decreasing the fluid viscosity, we reduce the influence of the fluid on the surface and find a convergence toward the analytical solution of a pure Maxwell fluid surface.
This demonstrates not only that the proposed viscoelastic surface model coincides with a Maxwell model in flat space, but also validates the numerical discretization and implementation.

\subsection{Viscoelastic surface in shear flow}
\label{sec:surfaceInShearFlow}
In the following we consider a viscoelastic surface in shear flow. 
Any laminar flow can be decomposed into a rotational part and a shear part. The latter leads to a deformation of suspended objects by a complex interplay between the rheology of the object and hydrodynamic forces.
The prediction of shape changes and flow dynamics is of fundamental relevance to understand the biophysics of biological cells in a viscous fluid. Hundreds of papers have addressed this issue, see e.g. \cite{Abreu2014, Bacher2019, Yazdani2013}. Most of them use lipid vesicles (no cell cortex) or shells (purely elastic cortex) as a model system for various cell types, which is a quite a rough  simplification in many flow regimes. 

Given the developed viscoelastic surface model, we are in a position to present the first simulations which take the viscoelasticity of the cell cortex into account.
In the following, we investigate the transition from tank-treading to tumbling which has been observed for vesicles in shear flow, in simulations \cite{Tumbling1} and experiments \cite{Tumbling2}.

\begin{figure}[h]
    \centering
    \begin{subfigure}{0.18\textwidth}
    \centering
        \includegraphics[width=\textwidth]{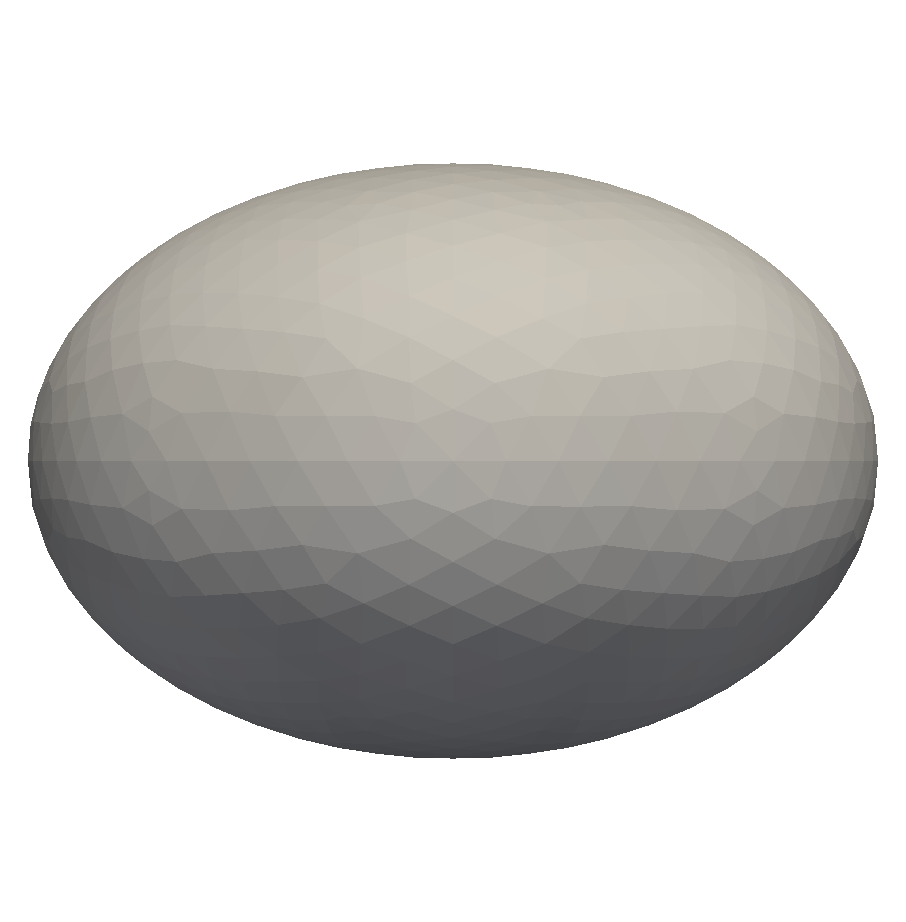}
        \caption{t=0}
        \label{fig:tumbling0}
    \end{subfigure}
    \begin{subfigure}{0.18\textwidth}
    \centering
        \includegraphics[width=\textwidth]{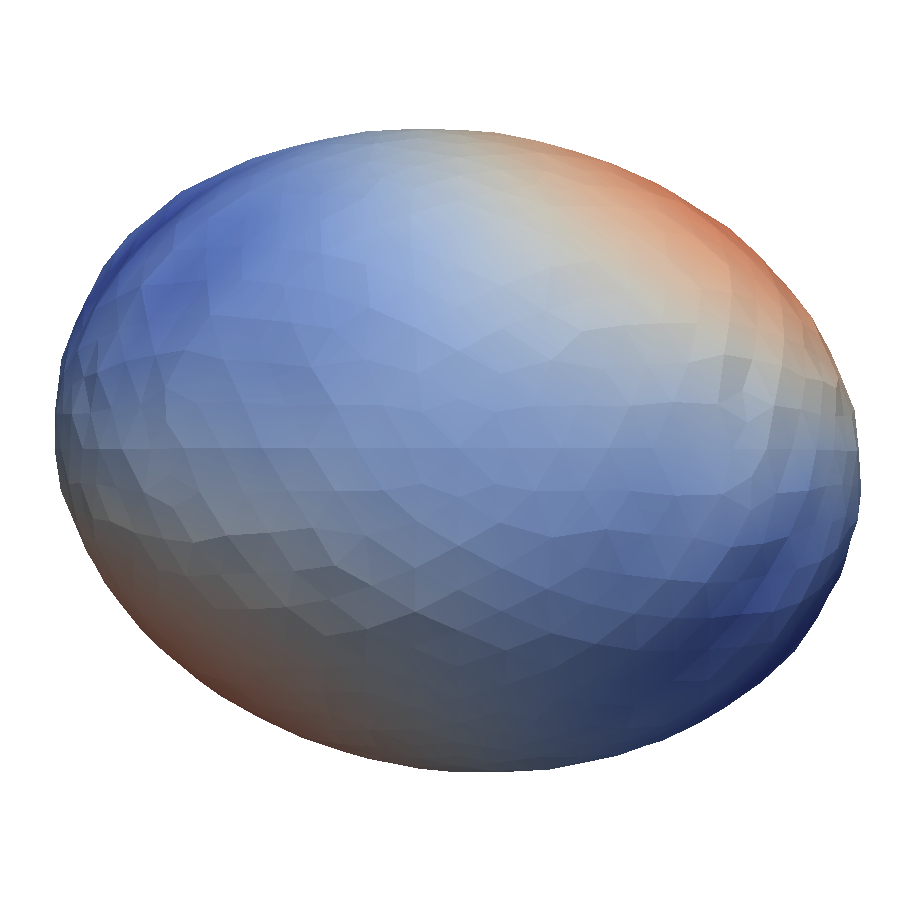}
        \caption{t=1.5}
        \label{fig:tumbling1}
    \end{subfigure}
    \begin{subfigure}{0.18\textwidth}
    \centering
        \includegraphics[width=\textwidth]{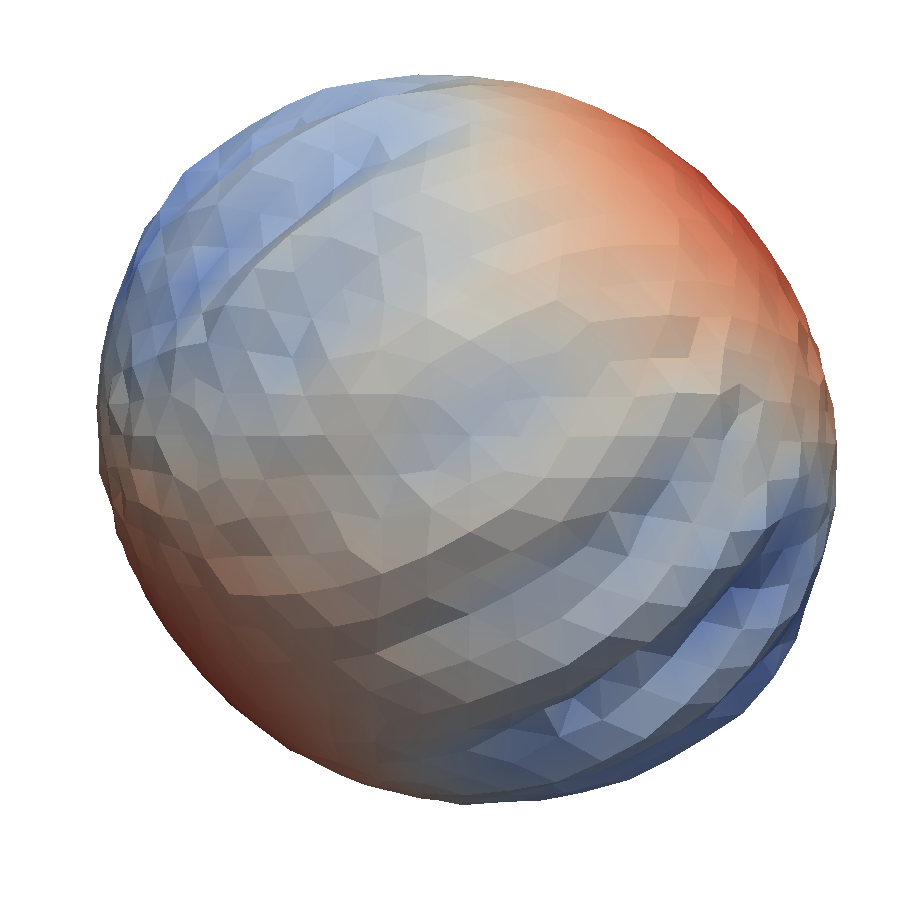}
        \caption{t=3.0}
        \label{fig:tumbling2}
    \end{subfigure}
    \begin{subfigure}{0.18\textwidth}
    \centering
        \includegraphics[width=\textwidth]{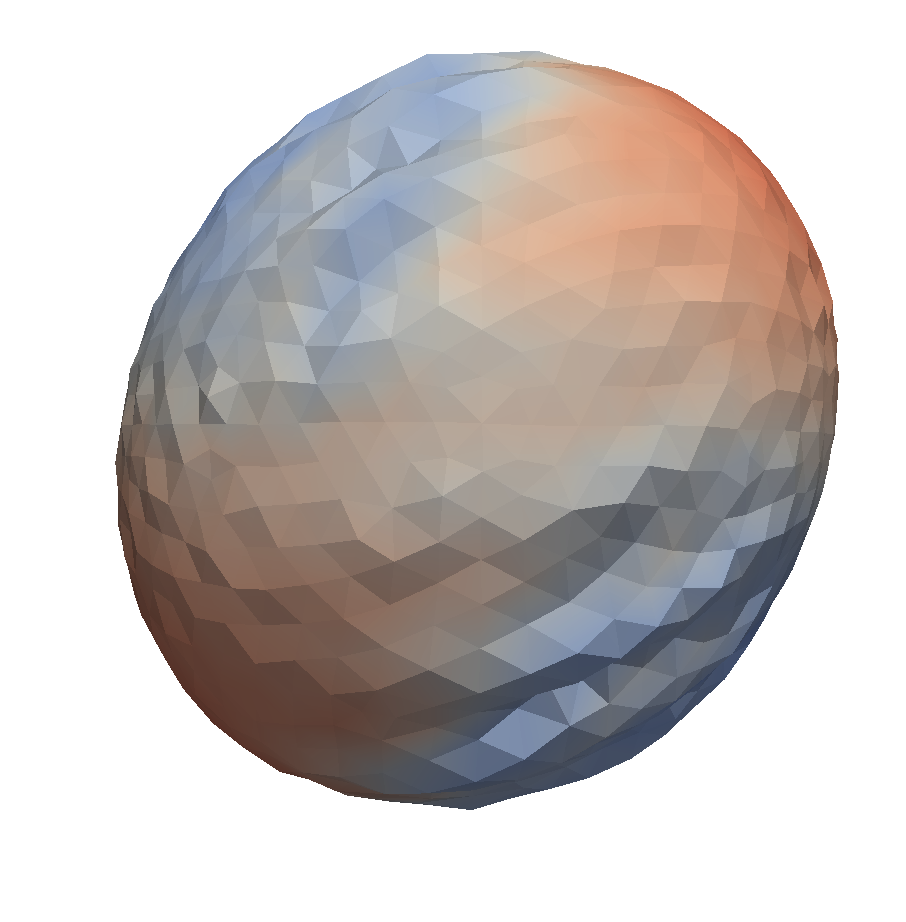}
        \caption{ t=4.5}
        \label{fig:tumbling3}
    \end{subfigure}
    \begin{subfigure}{0.22\textwidth}
    \centering
        \includegraphics[width=\textwidth]{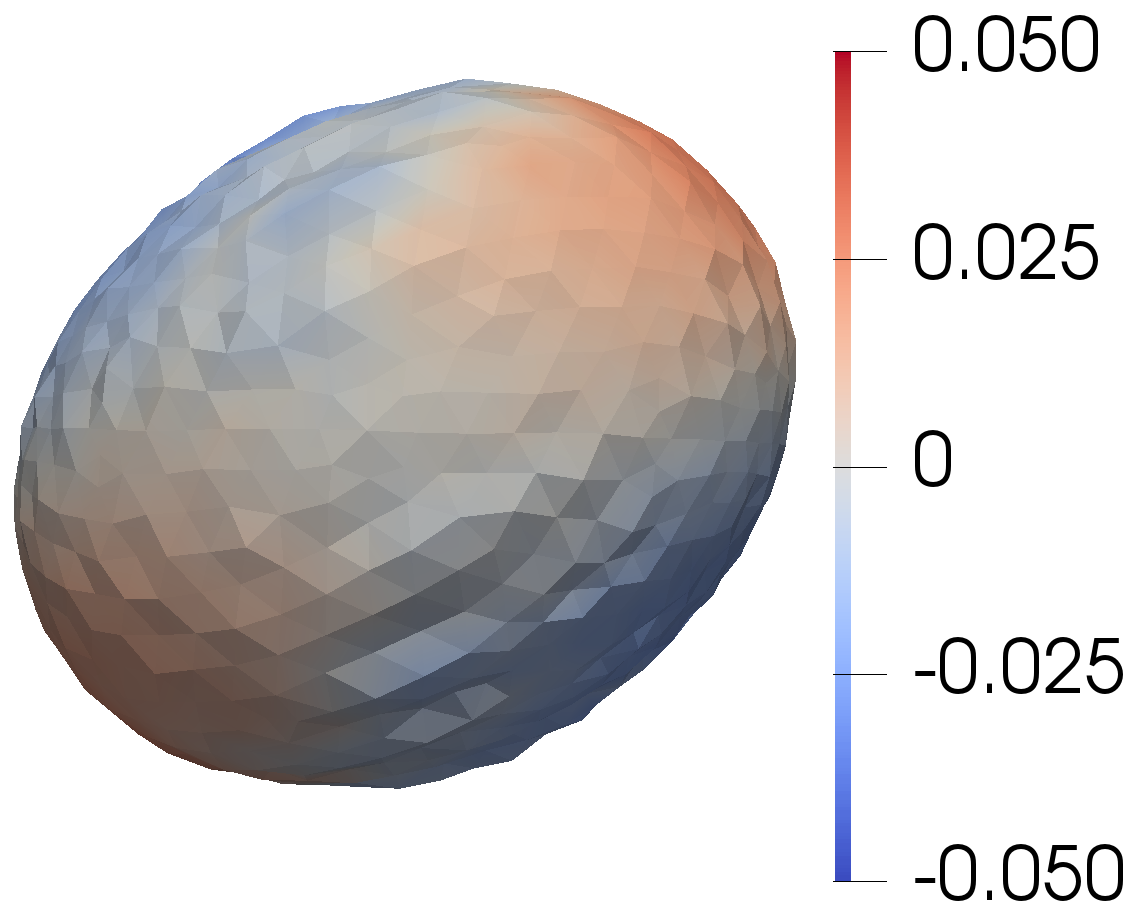}
        \caption{t=6.0}
        \label{fig:tumbling4}
    \end{subfigure}
    \caption{Tumbling, the parameter values are $\varepsilon_S = \varepsilon_A = 100$, $\gamma = 2$, the remaining parameters are 1. The colours represent $\tr S$, blue for negative, red for positive stress. Wiggles occur in the blue regions due to local compression of the surface.}
    \label{fig:tumblingSim}
\end{figure}

\begin{figure}[h]
    \centering
    \begin{subfigure}{0.182\textwidth}
    \centering
        \includegraphics[width=\textwidth]{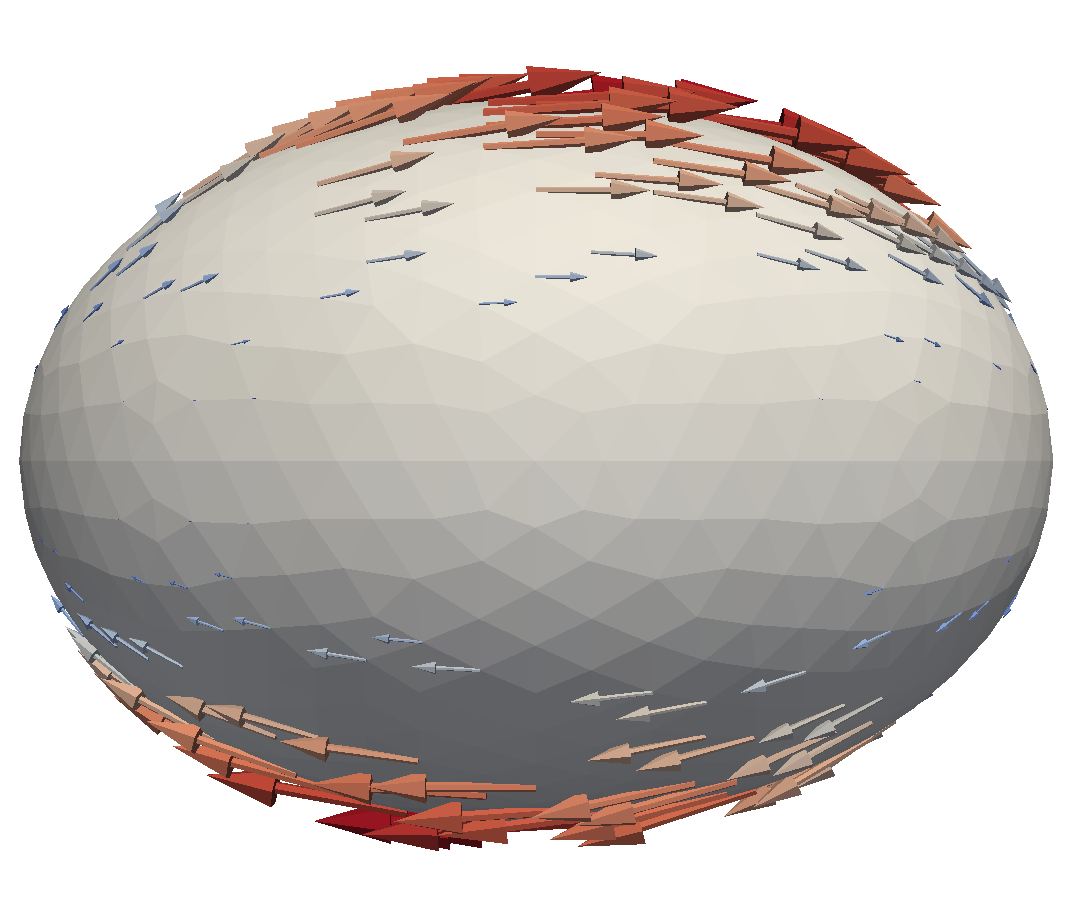}
        \caption{t=0}
        \label{fig:tanktreading0}
    \end{subfigure}
    \begin{subfigure}{0.182\textwidth}
    \centering
        \includegraphics[width=\textwidth]{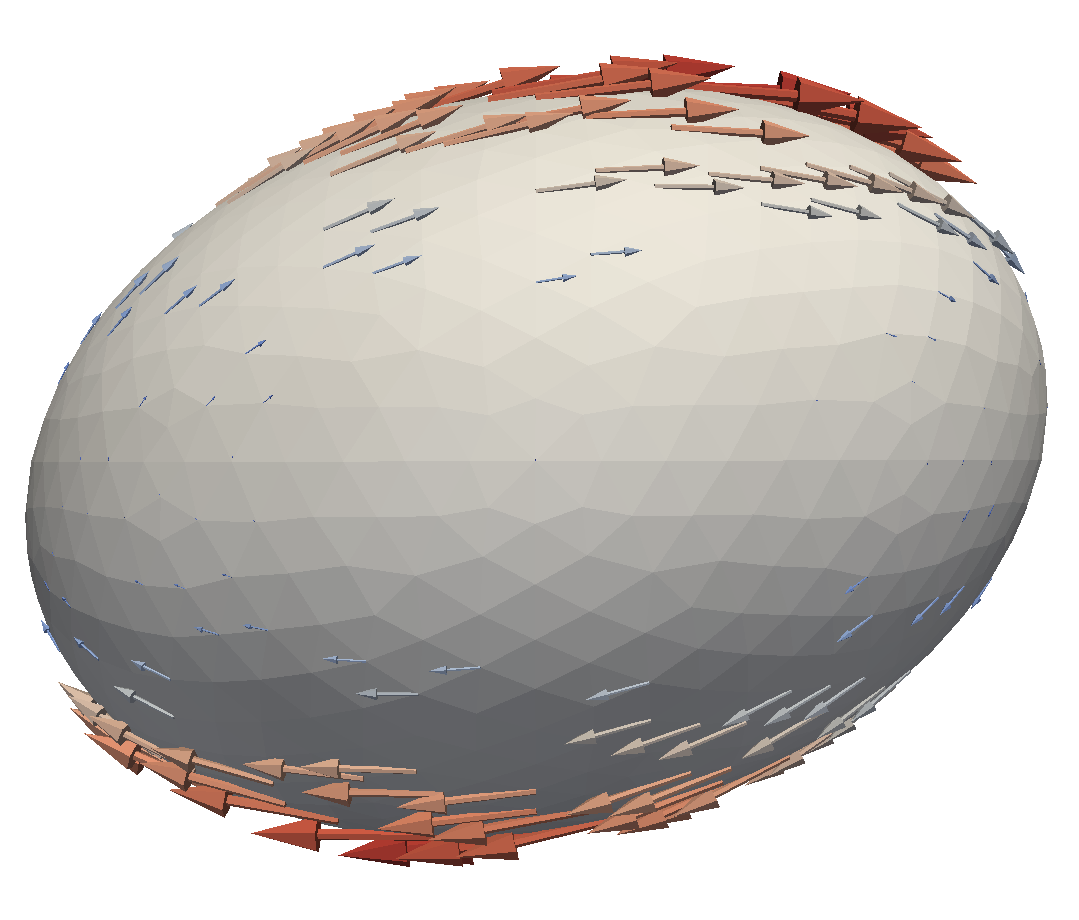}
        \caption{t=0.25}
        \label{fig:tanktread1}
    \end{subfigure}
    \begin{subfigure}{0.182\textwidth}
    \centering
        \includegraphics[width=\textwidth]{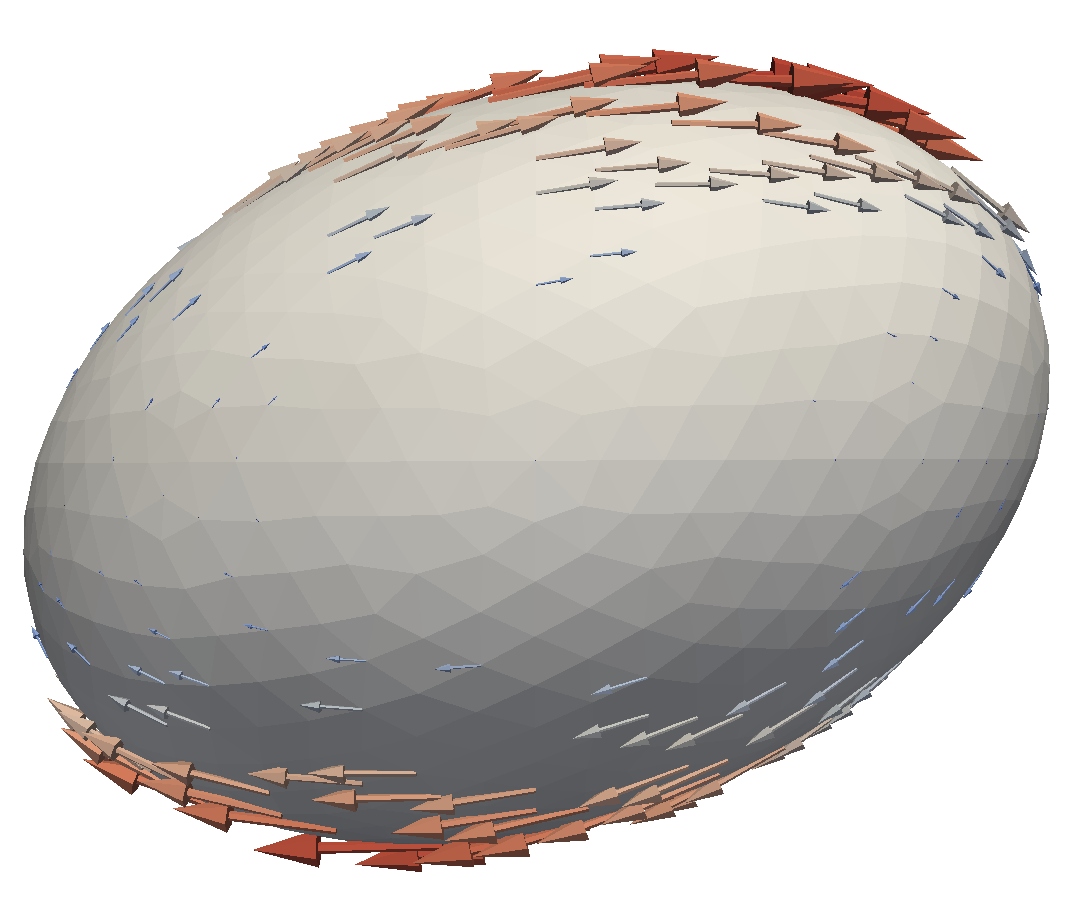}
        \caption{t=0.50}
        \label{fig:tanktreading2}
    \end{subfigure}
    \begin{subfigure}{0.182\textwidth}
    \centering
        \includegraphics[width=\textwidth]{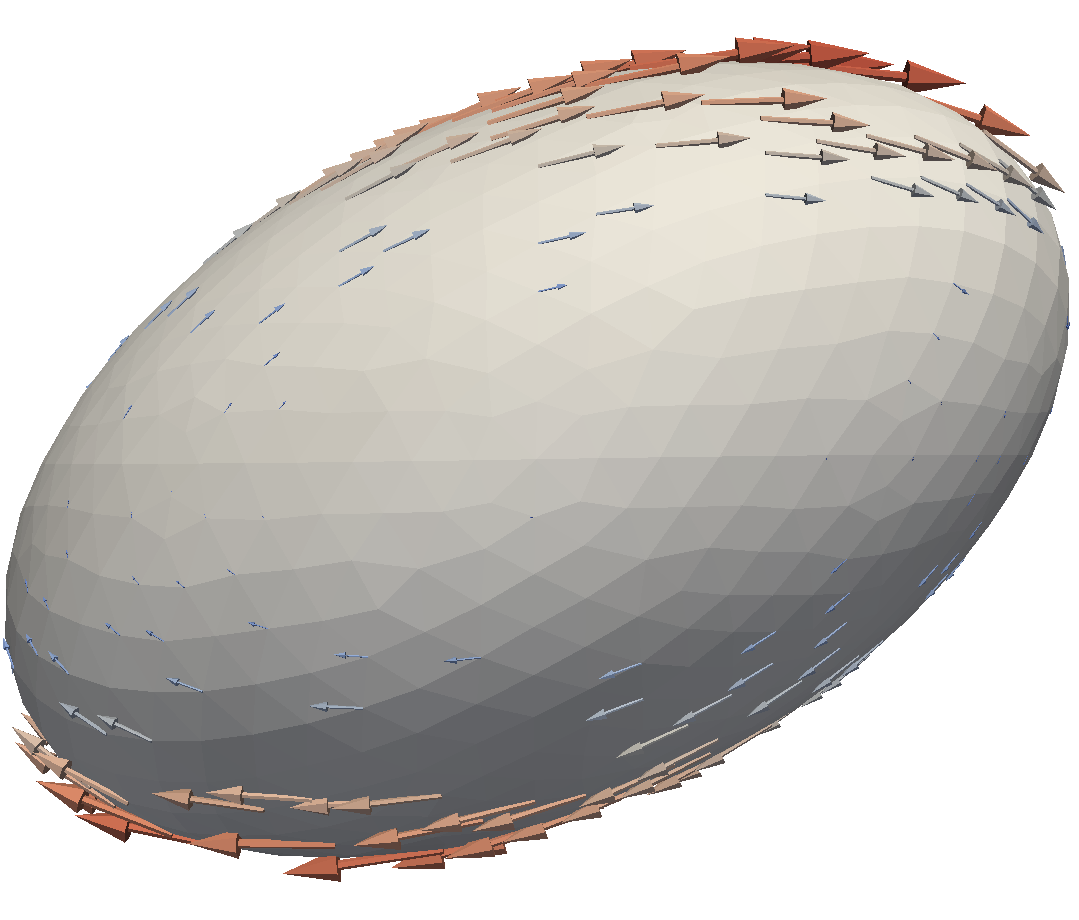}
        \caption{t=0.75}
        \label{fig:tanktreading3}
    \end{subfigure}
    \begin{subfigure}{0.22\textwidth}
    \centering
        \includegraphics[width=\textwidth]{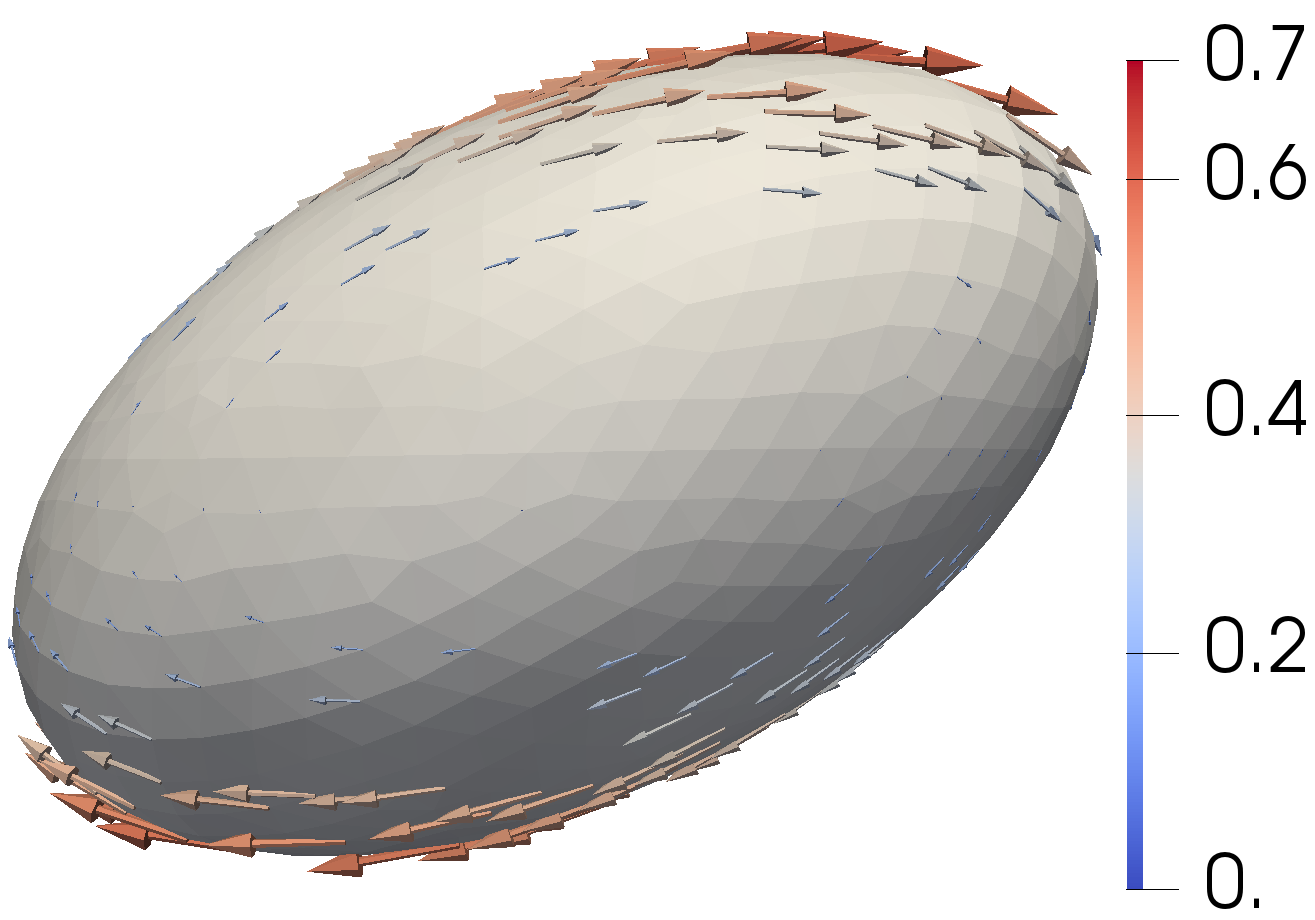}
        \caption{t=1.0}
        \label{fig:tanktreading4}
    \end{subfigure}
    \caption{Tank-treading, $\gamma = 2$, the remaining parameters are 1. The arrows show the tangential part of the flow field, illustrating the tank-treading behavior. The colour indicates the velocity.} 
    \label{fig:tanktreadingSim}
\end{figure}

The initial shape of the cell is prescribed as an ellipsoid (Figures \ref{fig:tumbling0} and \ref{fig:tanktreadingSim}) with two radii of length $1$ parallel to the $x$ and $z$-axes and one radius of length $0.7$, parallel to the $y$ axis. We impose a shear flow by setting the Dirichlet boundary condition,
\[
\vv{v} = U_0 \begin{pmatrix} y \\0 \\0 \end{pmatrix} \qquad \text{on } \partial \Omega_0 \setminus \partial \Omega_1
\]
on the outer fluid domain. 
For the parameters we restrict the simulations to $\eta_0=\eta_1=1$. As cellular flows are mostly in the low Reynolds number regime we eliminate inertial and time dependent terms by setting $\rho=\rho_\Gamma=0$. Cellular surface tension (typically created by Myosin motor proteins \cite{Fischer-Friedrich2014}) is set to $\gamma=2$. 
The viscoelastic parameters of the cell surface are varied to analyze their influence on the shape dynamics. 

\begin{figure}[H]
    \centering
    \sbox\threesubbox{%
      \resizebox{\dimexpr.9\textwidth-1em}{!}{%
        \includegraphics[height=3cm]{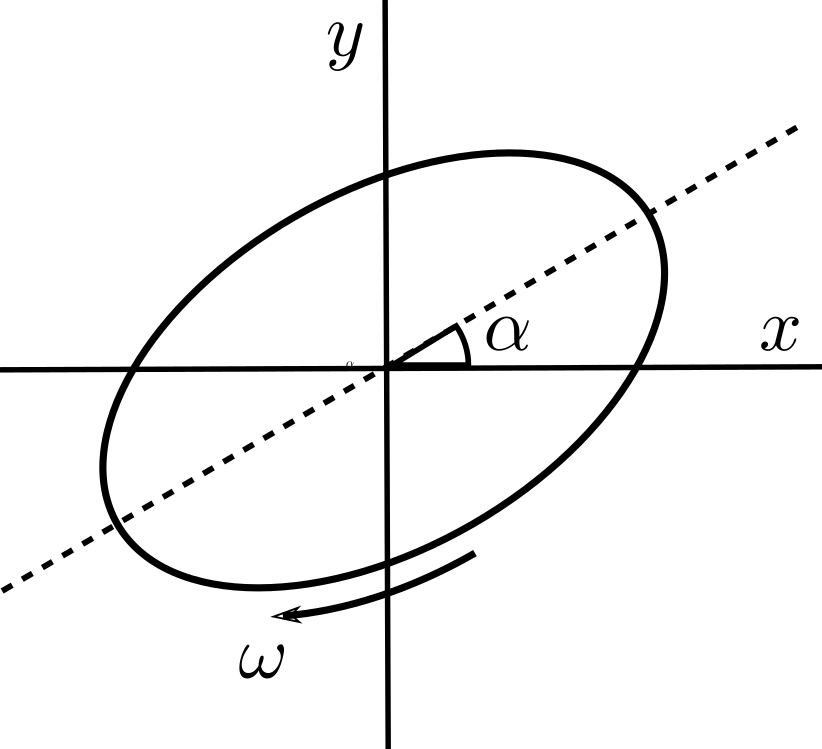}%
        \includegraphics[height=3cm]{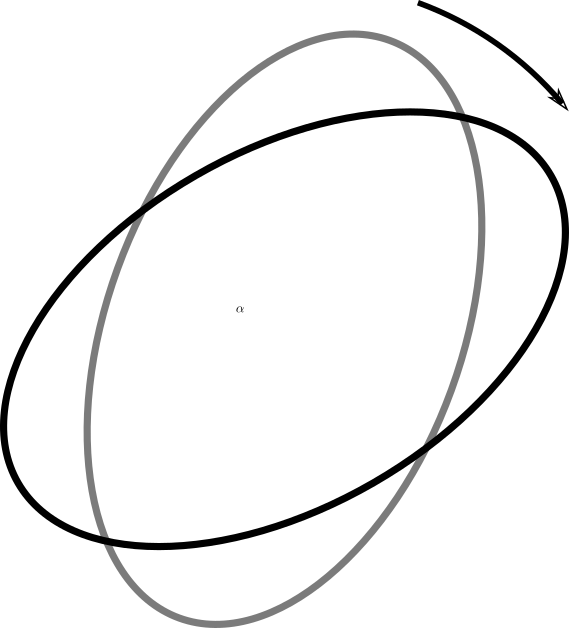}%
        \includegraphics[height=3cm]{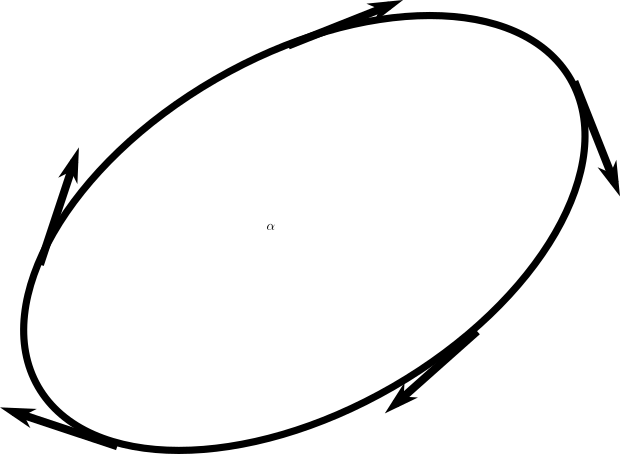}%
      }%
    }
    \setlength{\threesubht}{\ht\threesubbox}
    \centering
    \subcaptionbox{Inclination angle $\alpha$ and tank-treading frequency $\omega$.\label{fig:IncAngleSketch}}{%
      \includegraphics[height=0.7\threesubht]{Numerical results/InclinationAngleSketch.png}%
    }\qquad
    \subcaptionbox{Tumbling.\label{fig:tumbling}}{%
      \includegraphics[height=0.7\threesubht]{Numerical results/Tumbling.png}%
    }~~~~
    \subcaptionbox{Tank-treading.\label{fig:tanktreading}}{%
      \includegraphics[height=0.6\threesubht]{Numerical results/TankTreadingSketch.png}%
    }\qquad
    \caption{Schematic explanation of the inclination angle $\alpha$, tank-treading frequency $\omega$, tumbling and tank-treading.}
    \label{fig:schematic}
\end{figure}

\begin{figure}[H]
    \centering
   \includegraphics[width = \textwidth]{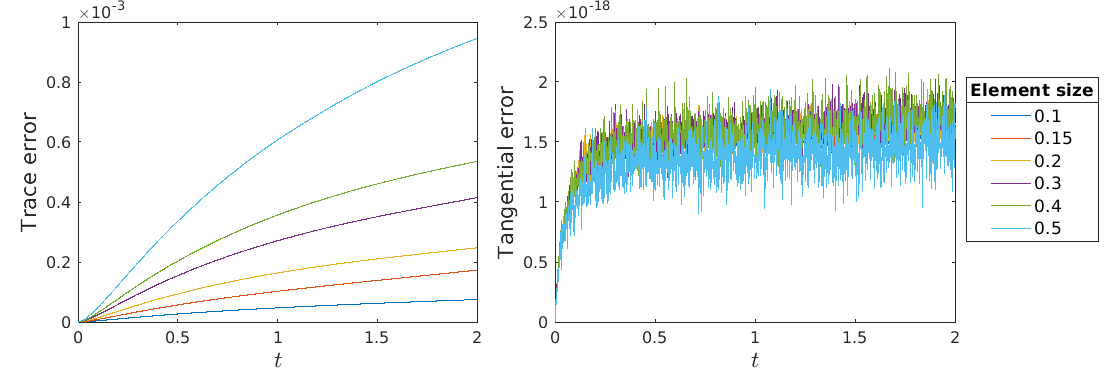}
    \caption{Measures for the numerical errors w.r.t. time $t$ for the tumbling case shown in Figure \ref{fig:tumblingSim}, for different element sizes on the surface. The trace error is defined as $\frac{1}{|\Gamma|}\int_\Gamma \tr \bar{S} d\Gamma$, the tangential error is defined as $ \frac{1}{|\Gamma|} \int_\Gamma ||\bar{S} \cdot n||_2 d\Gamma$.}
    \label{fig:errorMeasures}
\end{figure}

Assuming that the shape of the cell remains similar to an ellipsoid during the simulation, we can express its orientation by the inclination angle $\alpha$ between the long radius of the ellipse in the $x$,$y$-plane and the $x$-axis (Figure \ref{fig:IncAngleSketch}). 
Using the inclination angle $\alpha$ we can categorize the solutions into three groups.
\begin{enumerate}
    \item When the surface stress is low, there is little resistance against surface material points moving w.r.t. each other. So the cell can assume a stationary position in the flow plane (a fixed inclination angle $\alpha$) while its surface rotates around the interior fluid in tangential direction. This phenomenon is called tank-treading. The frequency of this periodic movement around the $z$-axis is the tank-treading frequency $\omega$ (Figure \ref{fig:IncAngleSketch} and \ref{fig:tanktreading}).
    \item  When there is more resistance to the surface material points moving w.r.t. each other, the surface will start rotating around the $z$-axis as a whole, similar to a rigid body, this is called tumbling (Figure \ref{fig:tumbling}).
    \item Between tank-treading and tumbling there is a transitional state, first observed experimentally in \cite{Tumbling2}. This state is characterized by irregular oscillations of the cell orientation.
    The ellipsoid moves but does not undergo a full rotation. 
\end{enumerate}
An numerical simulation  example of these states is given in Figure \ref{fig:tumblingSim} for tumbling and Figure \ref{fig:tanktreadingSim} for tank-treading. \\
As a measure for the numerical error of a simulation we use that the deviatoric part of the stress $\bar{S}$ must be tangential to $\Gamma$ and its trace $\tr \bar{S}$ must be zero. So to measure the error we calculate $ \frac{1}{|\Gamma|} \int_\Gamma ||\bar{S} \cdot n||_2 d\Gamma$ and $\frac{1}{|\Gamma|} \int_\Gamma \tr \bar{S} d \Gamma$. In Figure \ref{fig:errorMeasures} we show the error measures of a simulation with the same parameters as the simulation in Figure \ref{fig:tumblingSim} as a demonstration. For decreasing the element size of the surface mesh $T_\Gamma$, the error measure $\frac{1}{|\Gamma|} \int_\Gamma \tr \bar{S} d \Gamma$ decreases. The tangential error measure is negligible for any element size. Note that these measurements were done, without making $\bar{S}$ traceless at the end of each iteration, as was discussed at the end of Section \ref{sec:spatial discretization}.

\begin{figure}[h]
    \centering
    \begin{subfigure}{0.49\textwidth}
    \centering
        \includegraphics[width=0.99\textwidth]{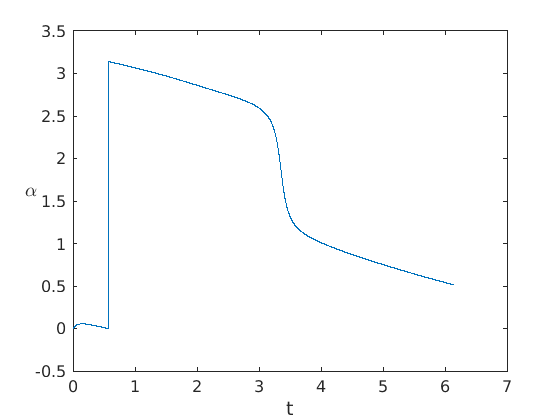}
        \caption{Tumbling}
        \label{fig:tumblingSimGraph}
    \end{subfigure}
    \begin{subfigure}{0.49\textwidth}
    \centering
        \includegraphics[width=0.99\textwidth]{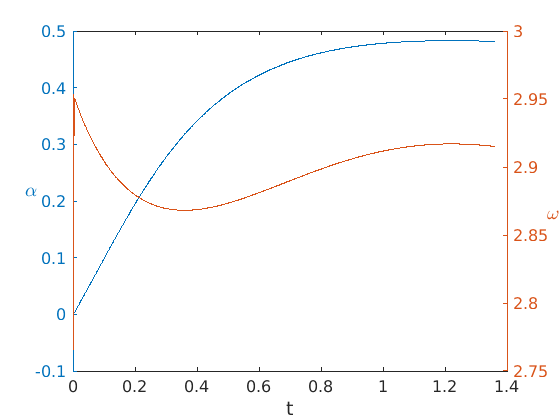}
        \caption{Tank-treading}
        \label{fig:tankTreadingSimGraph}
    \end{subfigure}
    \caption{Inclination angle $\alpha$ and tank-treading frequency $\omega$ w.r.t. time for the two example simulations in Figures \ref{fig:tumblingSim} and \ref{fig:tanktreadingSim}}
\end{figure}

In both scenarios the cells change shape. In the Tumbling case the cell rotates around the $z$-axis, its inclination angle is plotted in Figure \ref{fig:tumblingSimGraph} but it also tries to deform to a sphere because of the surface tension. When the difference between the long and short radii of the ellipsoid becomes too small, the tumbling will stop, as tumbling and tank-treading are equal for a sphere. The wrinkles that appear in the surface seem to be a physical property. Our conjecture is that compressing the surface in one direction causes it to expand perpendicularly. But instead of the cell becoming wider, the expansion causes wrinkles. 

When the surface is less resilient against deformation, the material points on it can move w.r.t. each other. As a result the cell as a whole and therefore its inclination angle $\alpha$, remain stationary after some time. The material points of the membrane however, circulate around the $z$-axis with angular velocity $\omega$, see Figure \ref{fig:tankTreadingSimGraph}. The cell will be stretched slowly by the constant forces of the fluid acting on it and the dissipation of stress.

\begin{figure}[h]
    \centering
    \includegraphics[width=0.7\textwidth]{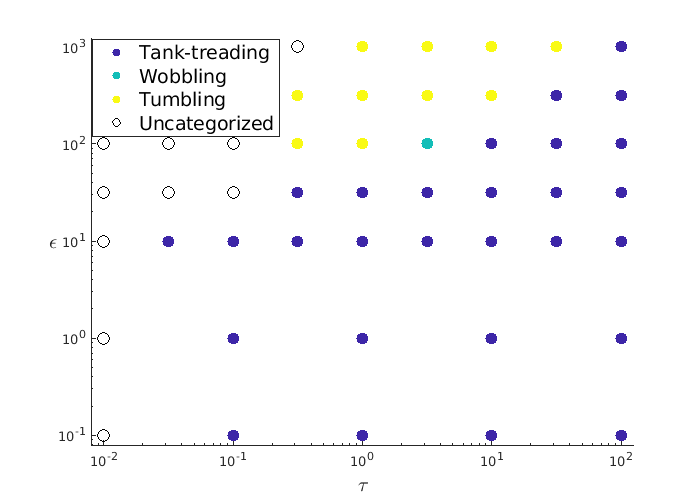}
    \caption{Phase diagram for the different solutions depending on $\tau_A=\tau_S=\tau$ and $\varepsilon_A = \varepsilon_S = \varepsilon$. $\gamma = 2$ and the remaining parameters are 1.}
    \label{fig:phaseDiagram}
\end{figure}

We tested the change in solution for changing the surface viscosity and the relaxation time. The results can be found in Figure \ref{fig:phaseDiagram}. We found that increasing the surface viscosity directly increases the surface stress, which results in a transition from tank-treading to tumbling. Increasing the relaxation time slowed the increase in surface stress, which eventually leads to a transition from tumbling to tank-treading. The transitional state, trembling, is just a temporary state. Due to the surface tension the long axis of the cell will shorten, changing trembling to tank-treading. The uncategorized simulations occur when $\varepsilon/\tau$ is too large, increasing the surface stress too rapidly. This resulted in a solution with a very distorted flow, which could not be categorized in these three categories.

\subsection{Outlook: Active cortex flow driving cell division}
\label{sec:outlook}
In the following we present another application to illustrate the versatility and applicability of our method for active cell membranes.
Here, we consider the onset of cytokinesis (the cellular deformation during cell division) triggered by active stresses in the cell cortex. As discussed in the introduction, the cell cortex is a prominent example of a viscoelastic surface. It surrounds most animal cells and can exert active contractile forces when additional motor proteins bind to it. 

Cell cytokinesis is a prominent subject of active matter research in biological physics. 
It is believed that the formation of a ring of high motor protein concentration at the cell equator drives the division of a cell into two daughter cells \cite{10.1083/jcb.201612068, grill2011growing, green2012cytokinesis}. 
Previous studies on cytokinesis have considered the cortical cell surface as purely viscous \cite{salbreux2009hydrodynamics,sedzinski2011polar,Mietke2018,Mietke2019}, purely elastic \cite{Bacher2019} or by a phase field approach without any in-plane rheology \cite{zhao2016modeling}. 

To simulate cytokinesis, we consider a dividing spherical cell, using the unit sphere centered at the origin as initial condition. The active tension from the motor proteins is explicitly incorporated by choosing a spatially dependent function for $\gamma(x)$,
\begin{equation}
\gamma(x) = \gamma_0
    \begin{cases}
    e^{-\left( \frac{x-\mu_1}{\sigma} \right)^2} + c &x < \mu_1 \\
    1 + c &x \in [\mu_1, \mu_2] \\
    e^{-\left( \frac{x-\mu_2}{\sigma} \right)^2} + c &x > \mu_2 \\
    \end{cases}.
\end{equation}
We chose the exponential function because it is smooth and decays rapidly for $x<\mu_1$ and $x>\mu_2$. Hence the surface tension is strongest in the ring around the cell with $x \in [\mu_1, \mu_2]$, mimicking the contractile ring observed in cytokinesis. We chose $\mu_1 = - \mu_2 = 0.2$ and $\sigma = 0.2$ to make the ring narrow and make the surface tension reduce quickly for $x \notin [\mu_1, \mu_2]$. We chose $c = 0.1$ to keep the surface smooth at the poles, and $\gamma_0 = 1$.

\begin{figure}[h]
    \centering
    \begin{subfigure}{0.32\textwidth}
    \centering
        \includegraphics[width=\textwidth]{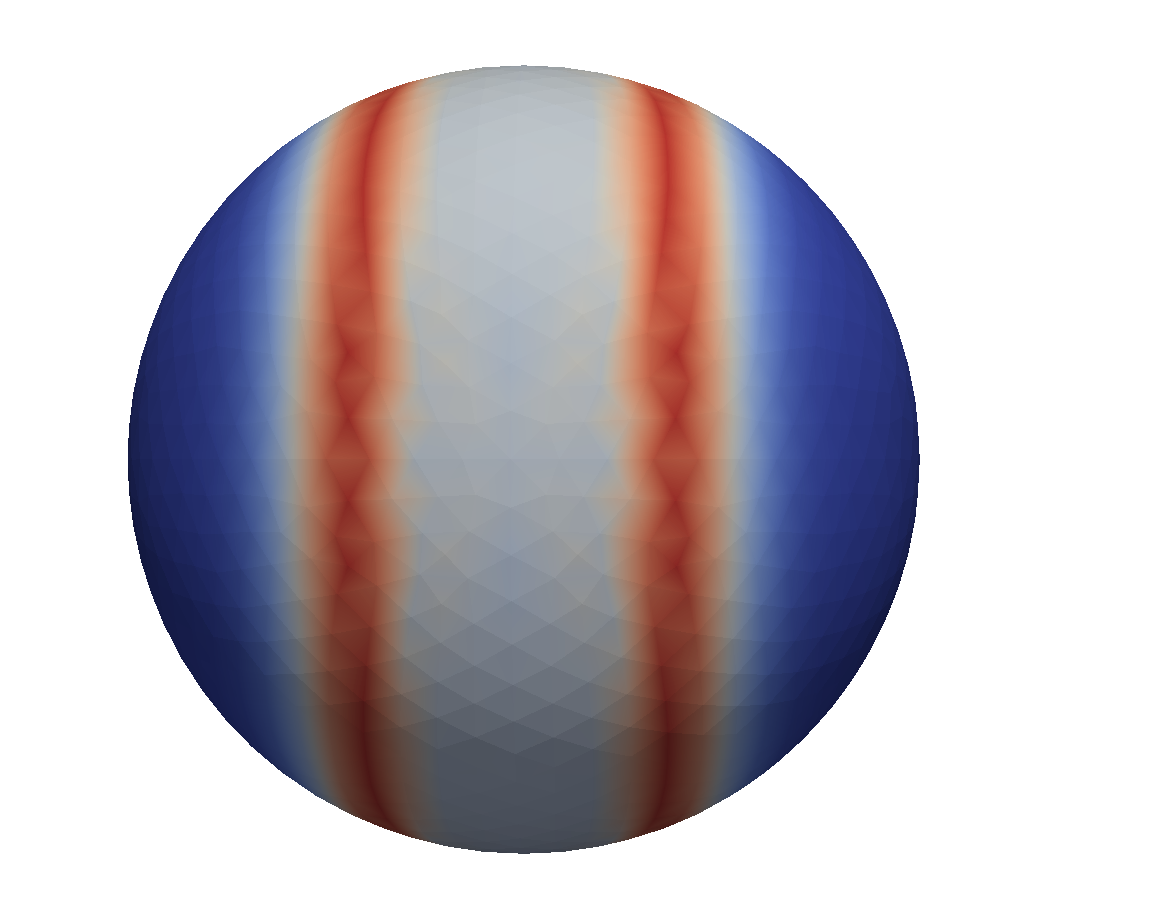}
        \caption{$t=0$}
        \label{fig:CellDivSurfTen2}
    \end{subfigure}
    \begin{subfigure}{0.32\textwidth}
    \centering
        \includegraphics[width=\textwidth]{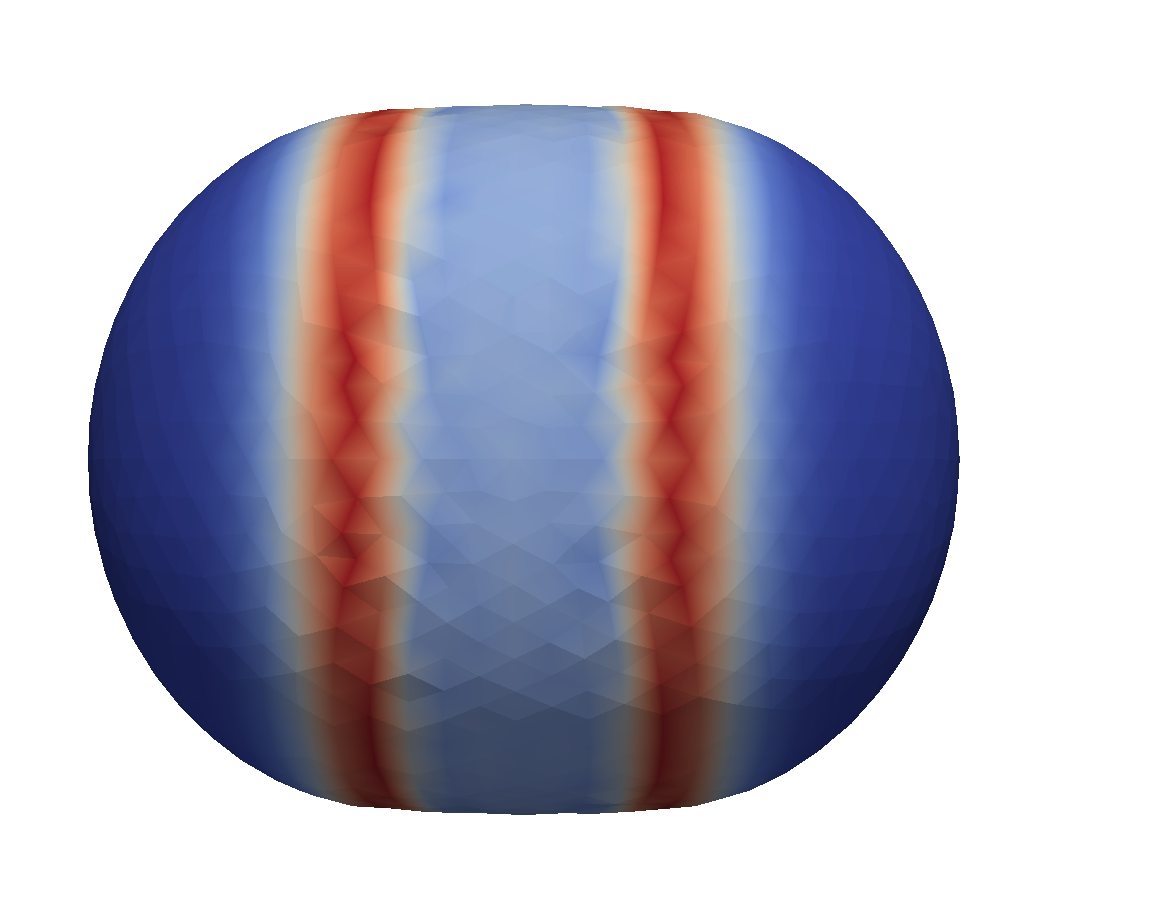}
        \caption{$t=3.5$}
        \label{fig:CellDivSurfTen19}
    \end{subfigure}
    \begin{subfigure}{0.32\textwidth}
    \centering
        \includegraphics[width=\textwidth]{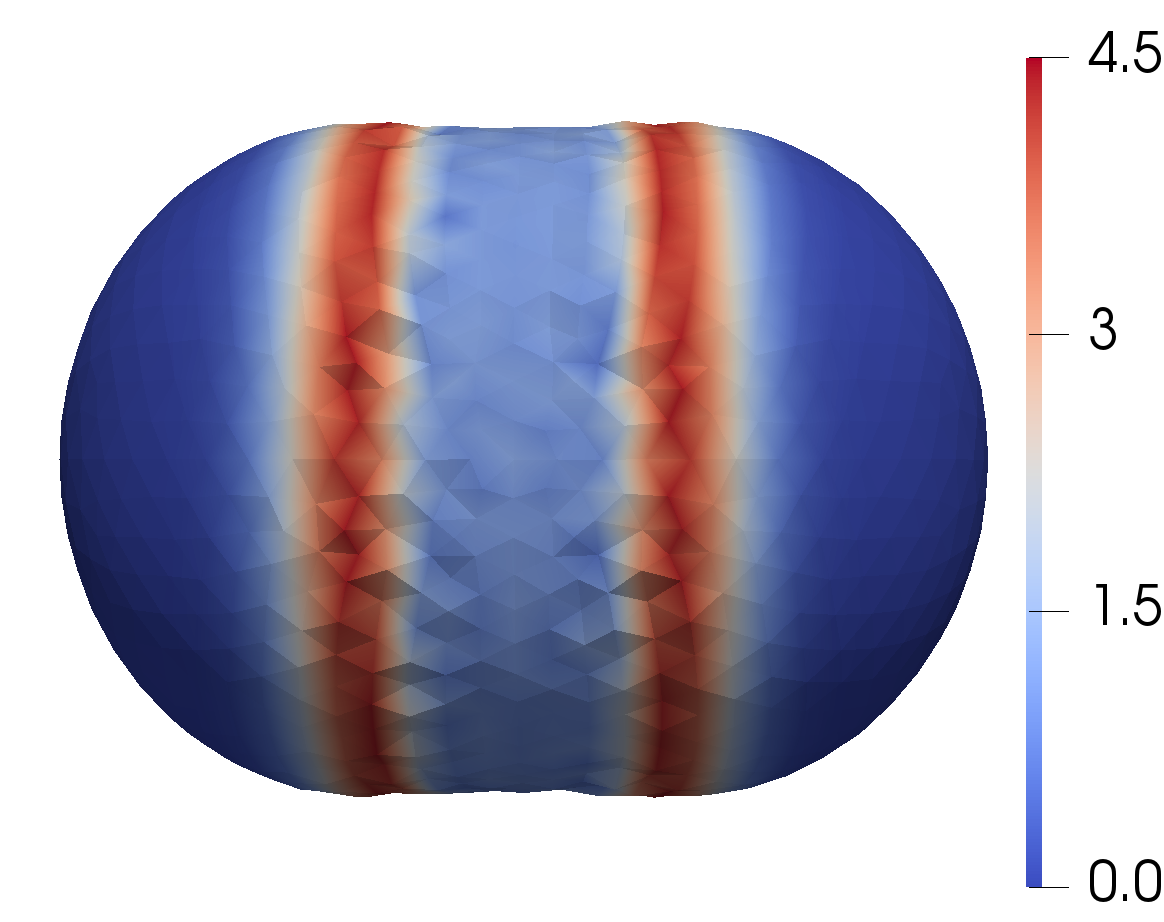}
        \caption{$t=7$}
        \label{fig:CellDivSurfTen37}
    \end{subfigure}
    
    \vspace{0.4cm}
    \begin{subfigure}{0.32\textwidth}
    \centering
        \includegraphics[width=\textwidth]{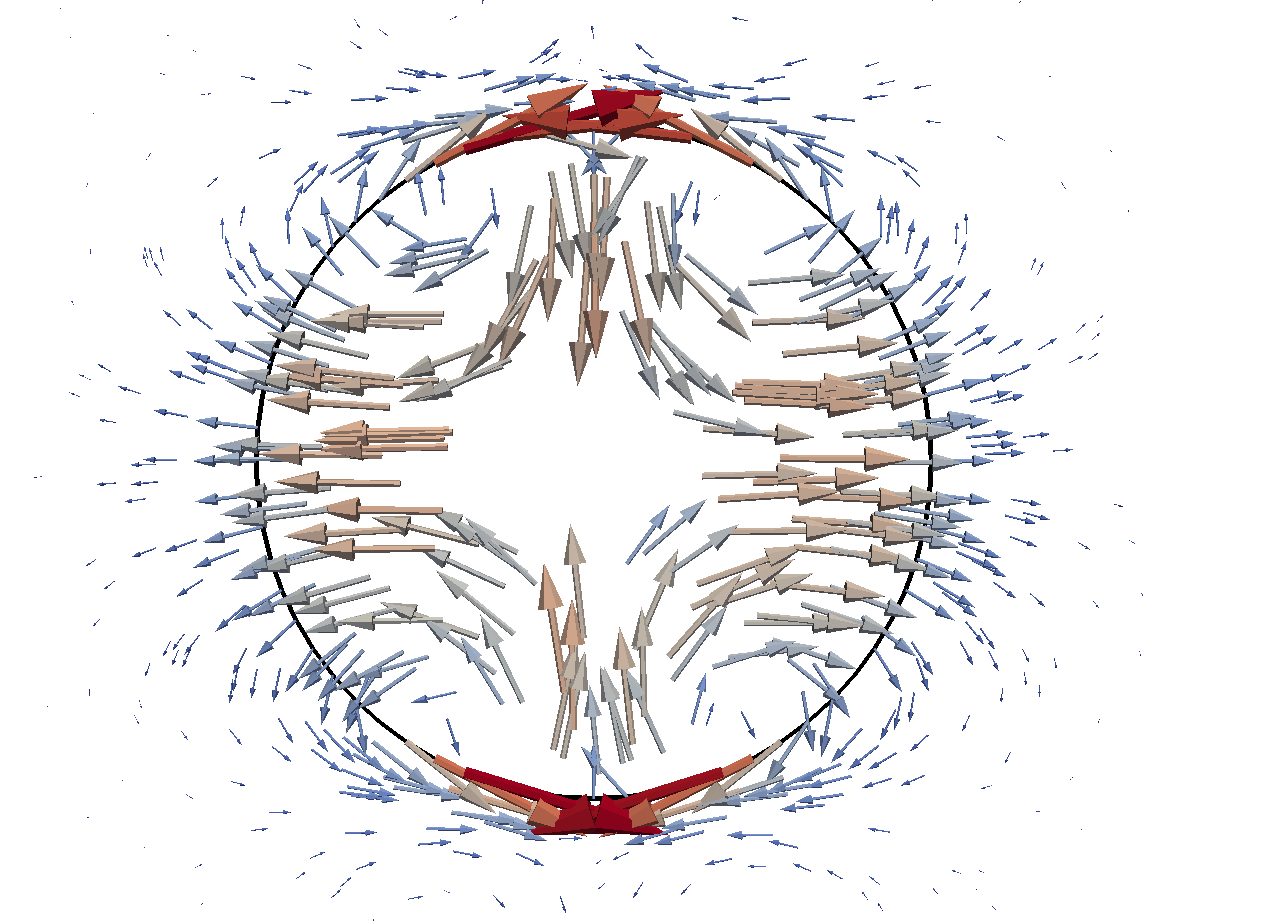}
        \caption{$t=0$}
        \label{fig:CellDivV2}
    \end{subfigure}
    \begin{subfigure}{0.32\textwidth}
    \centering
        \includegraphics[width=\textwidth]{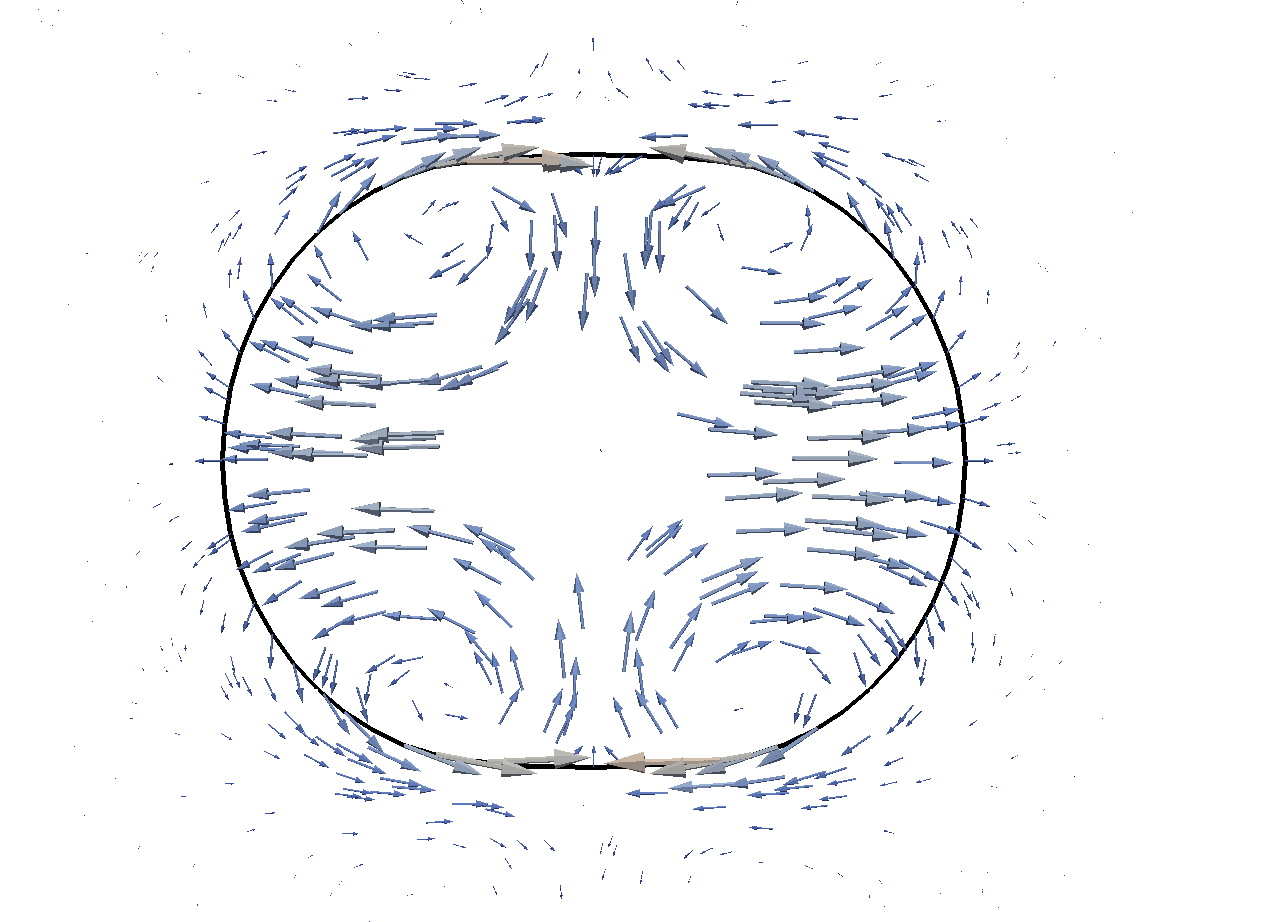}
        \caption{$t=3.5$}
        \label{fig:CellDivV19}
    \end{subfigure}
    \begin{subfigure}{0.32\textwidth}
    \centering
        \includegraphics[width=\textwidth]{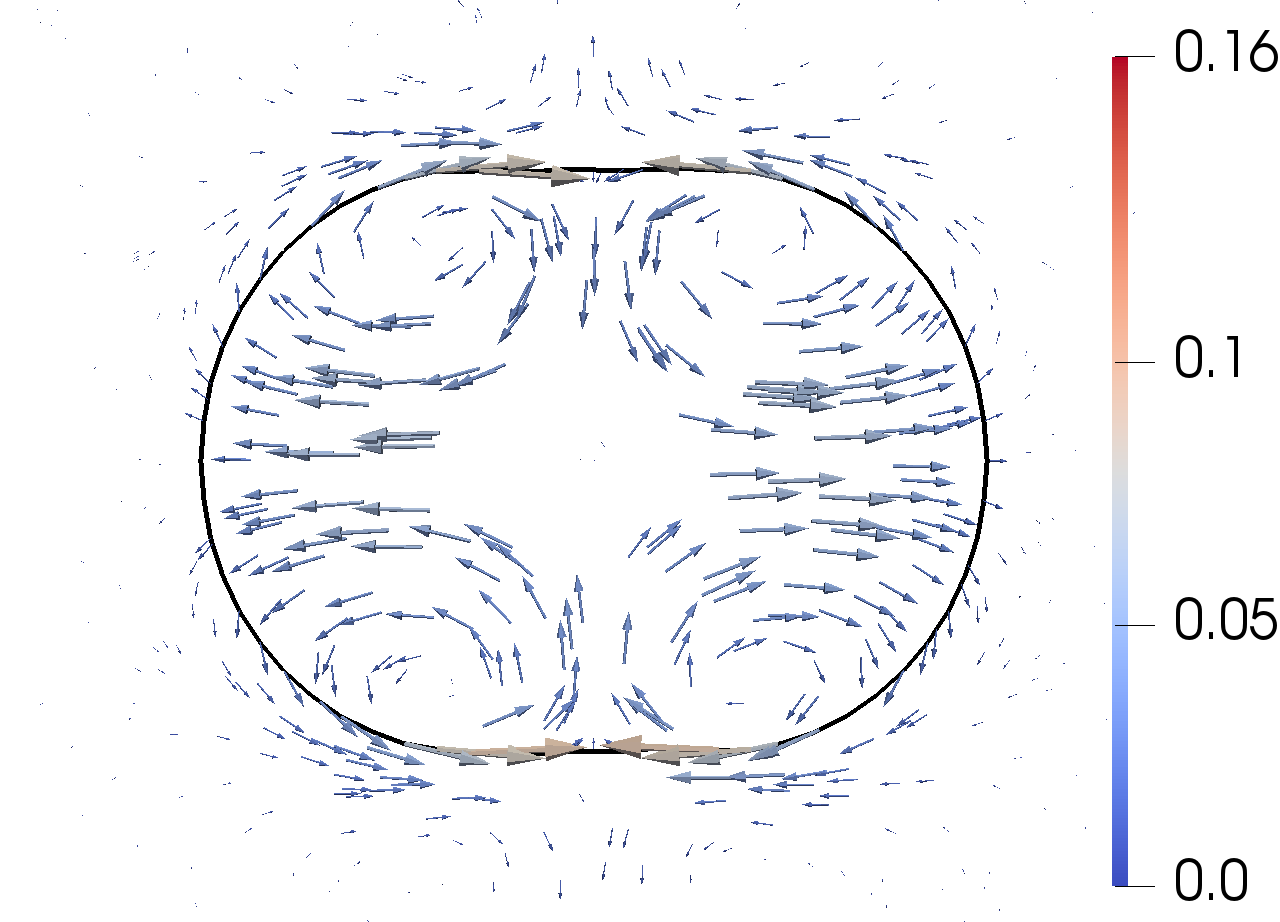}
        \caption{$t=7$}
        \label{fig:CellDivV37}
    \end{subfigure}
    
    \caption{Spherical vesicle during the onset of cytokinesis, with the force induced by the surface tension defined as, $\nabla_\Gamma \cdot (\gamma P)$, including the Marangoni contribution. 
    In (a)-(c) the color scaling represents the norm of the force generated by the surface tension.
    (d)-(f) shows the cross-section in the $x$,$y$-plane with cell surface in black and velocity vectors colored by their magnitude. 
    All parameter values are chosen equal to 1.}
    \label{fig:cellDivSurfTen}
\end{figure}

\begin{figure}[h]
    \centering
    \begin{subfigure}{0.32\textwidth}
    \centering
        \includegraphics[width=\textwidth]{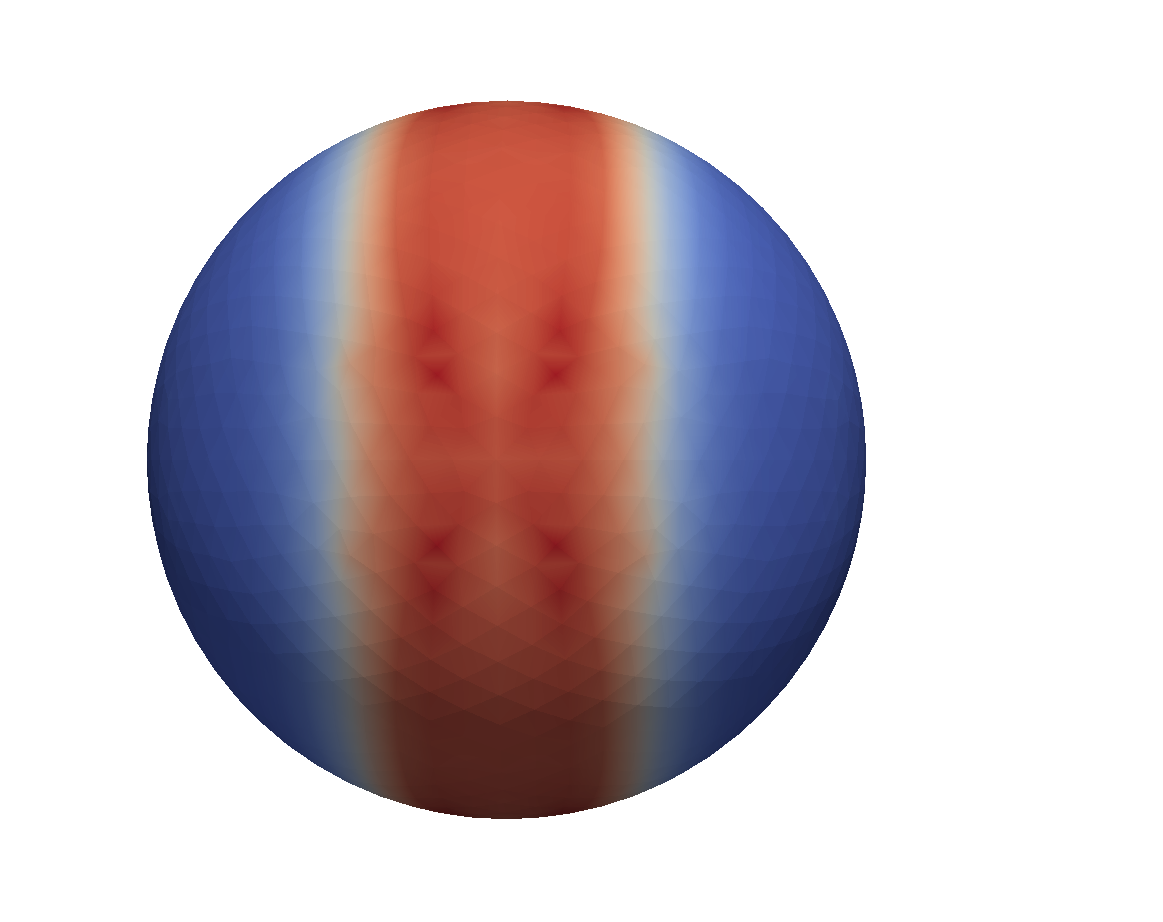}
        \caption{$t=0$}
        \label{fig:WOmarangoni2}
    \end{subfigure}
    \begin{subfigure}{0.32\textwidth}
    \centering
        \includegraphics[width=\textwidth]{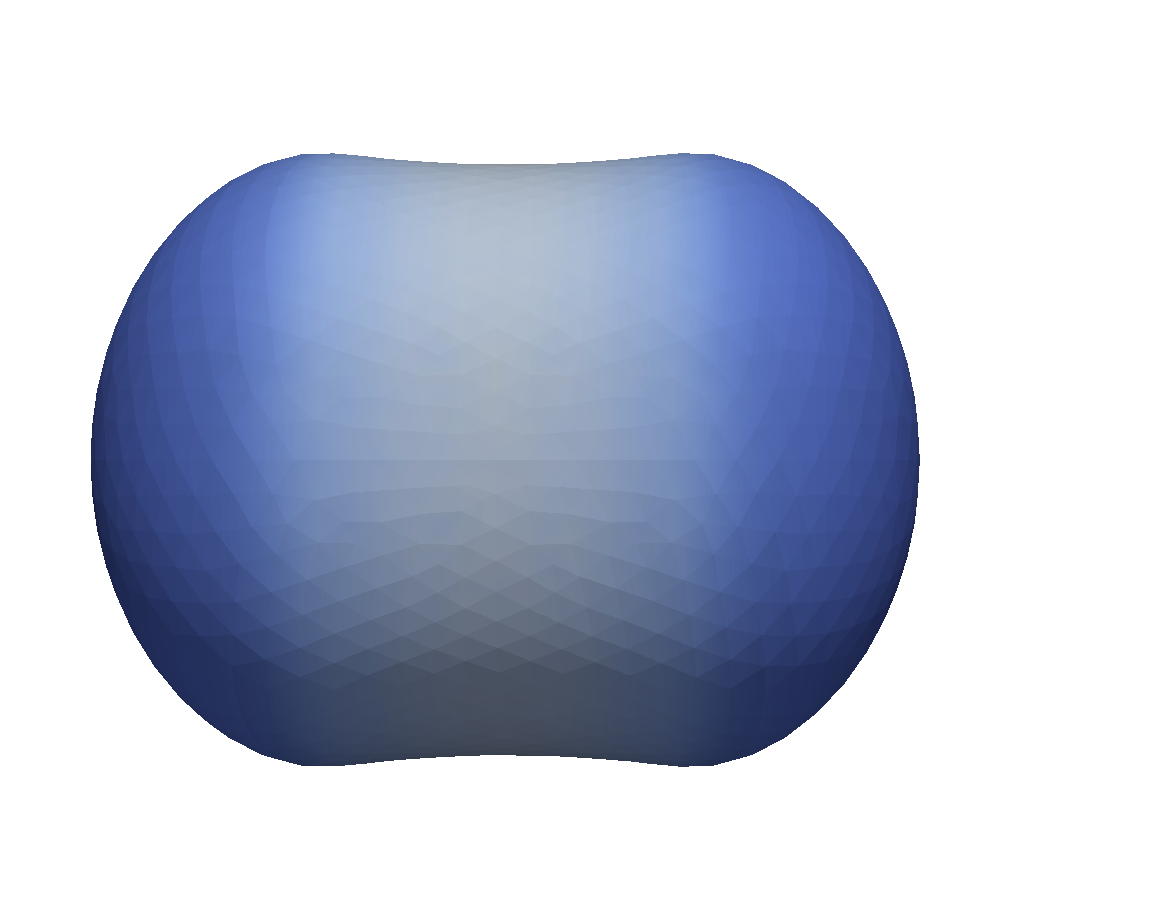}
        \caption{$t=3.5$}
        \label{fig:WOmarangoni19}
    \end{subfigure}
    \begin{subfigure}{0.32\textwidth}
    \centering
        \includegraphics[width=\textwidth]{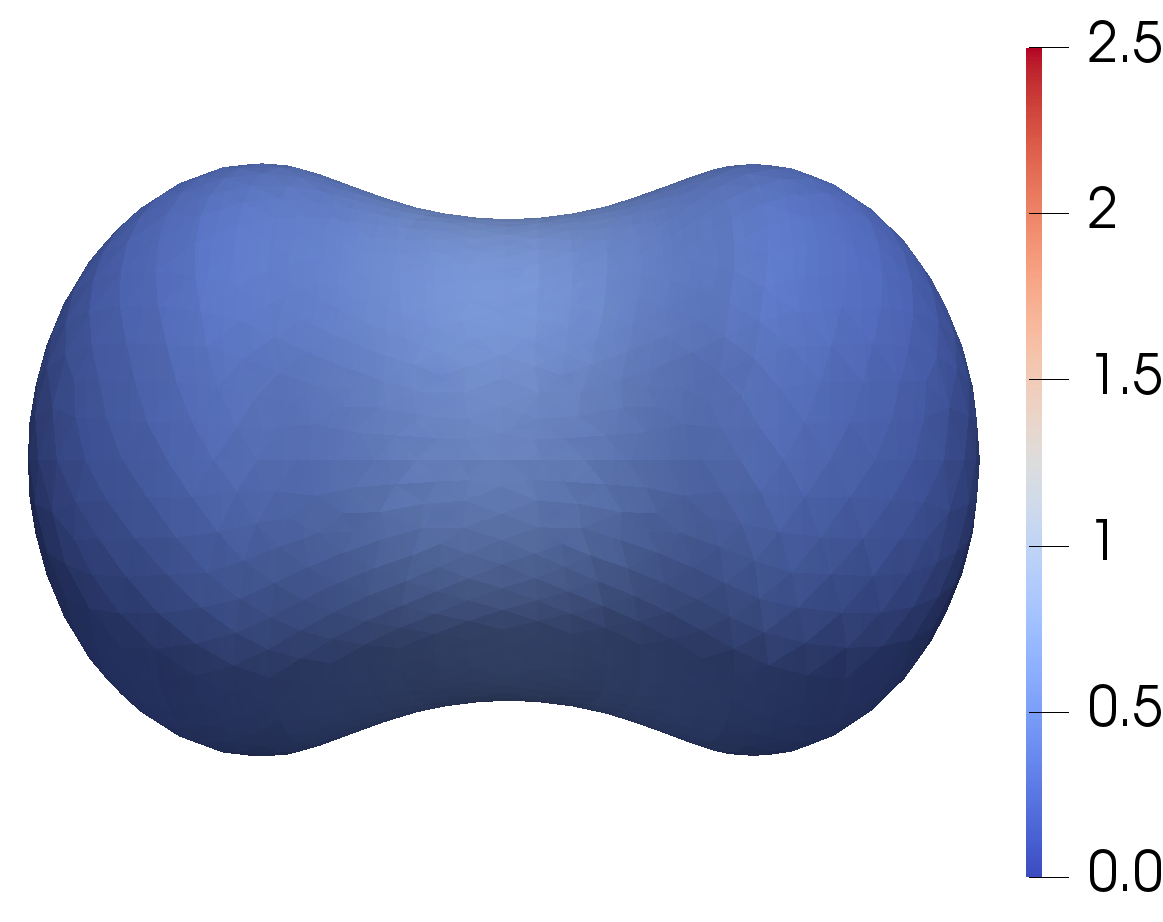}
        \caption{$t=14$}
        \label{fig:WOmarangoni37}
    \end{subfigure}
    
    \vspace{0.4cm}
    \begin{subfigure}{0.32\textwidth}
    \centering
        \includegraphics[width=\textwidth]{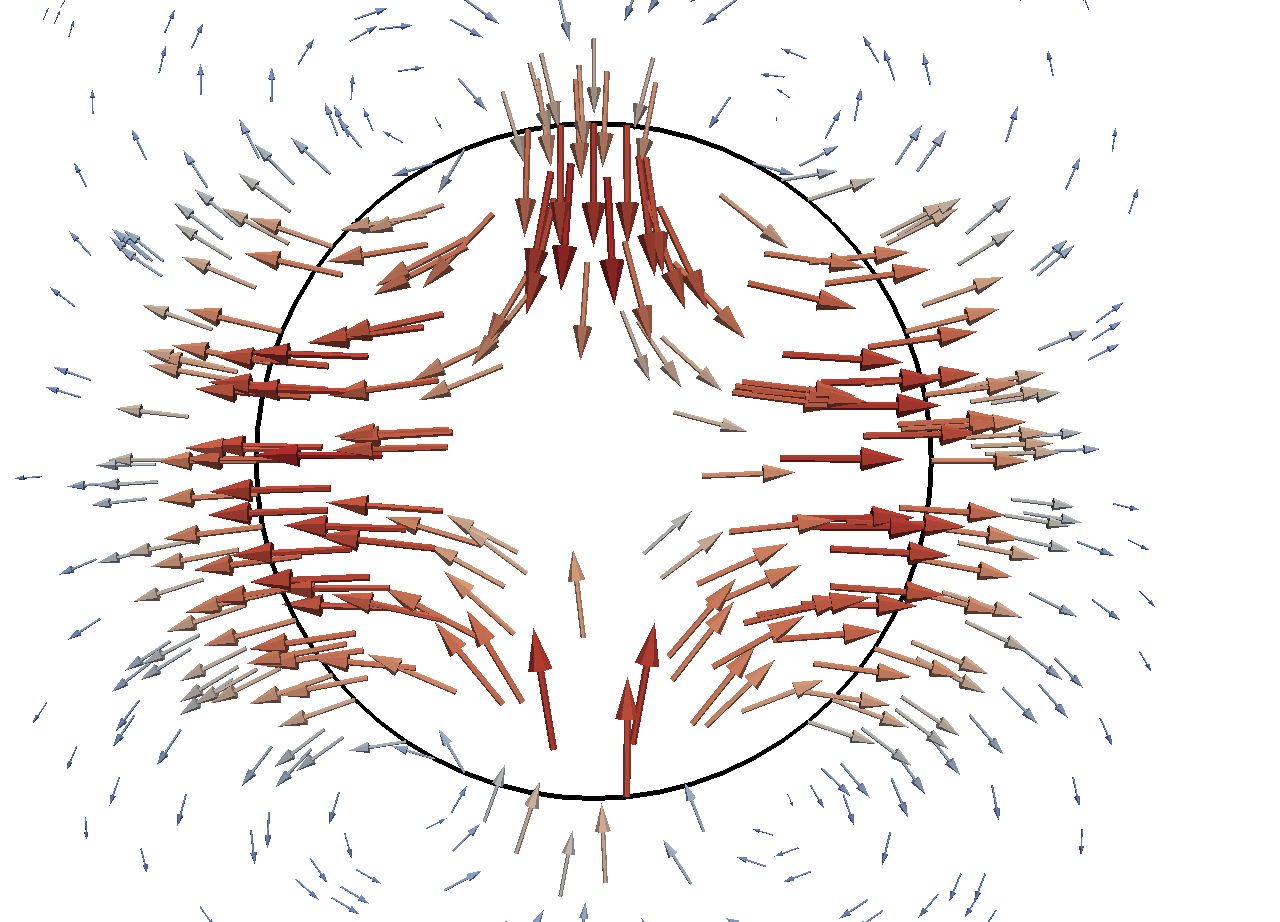}
        \caption{$t=0$}
        \label{fig:WOmarangoniV2}
    \end{subfigure}
    \begin{subfigure}{0.32\textwidth}
    \centering
        \includegraphics[width=\textwidth]{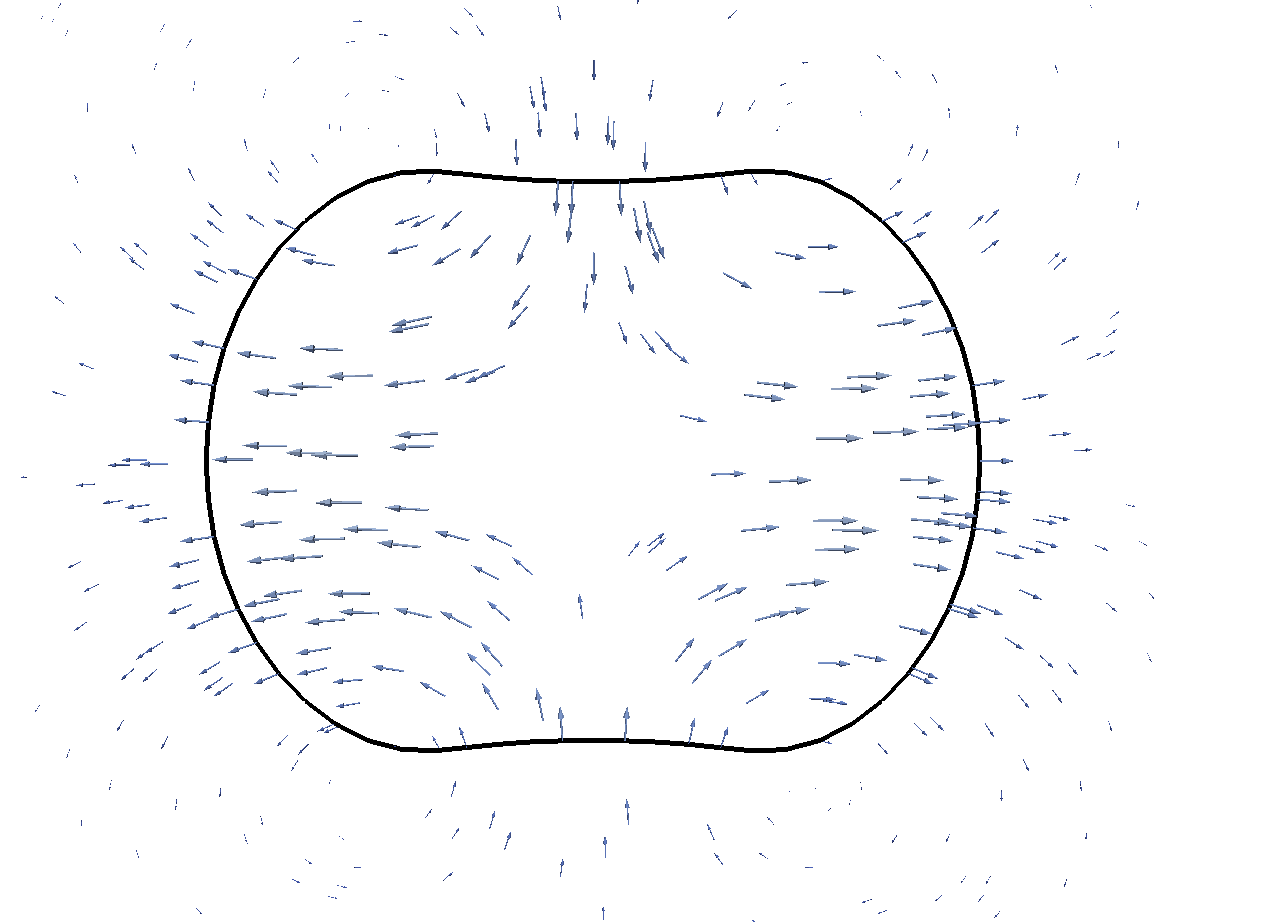}
        \caption{$t=3.5$}
        \label{fig:WOmarangoniV19}
    \end{subfigure}
    \begin{subfigure}{0.32\textwidth}
    \centering
        \includegraphics[width=\textwidth]{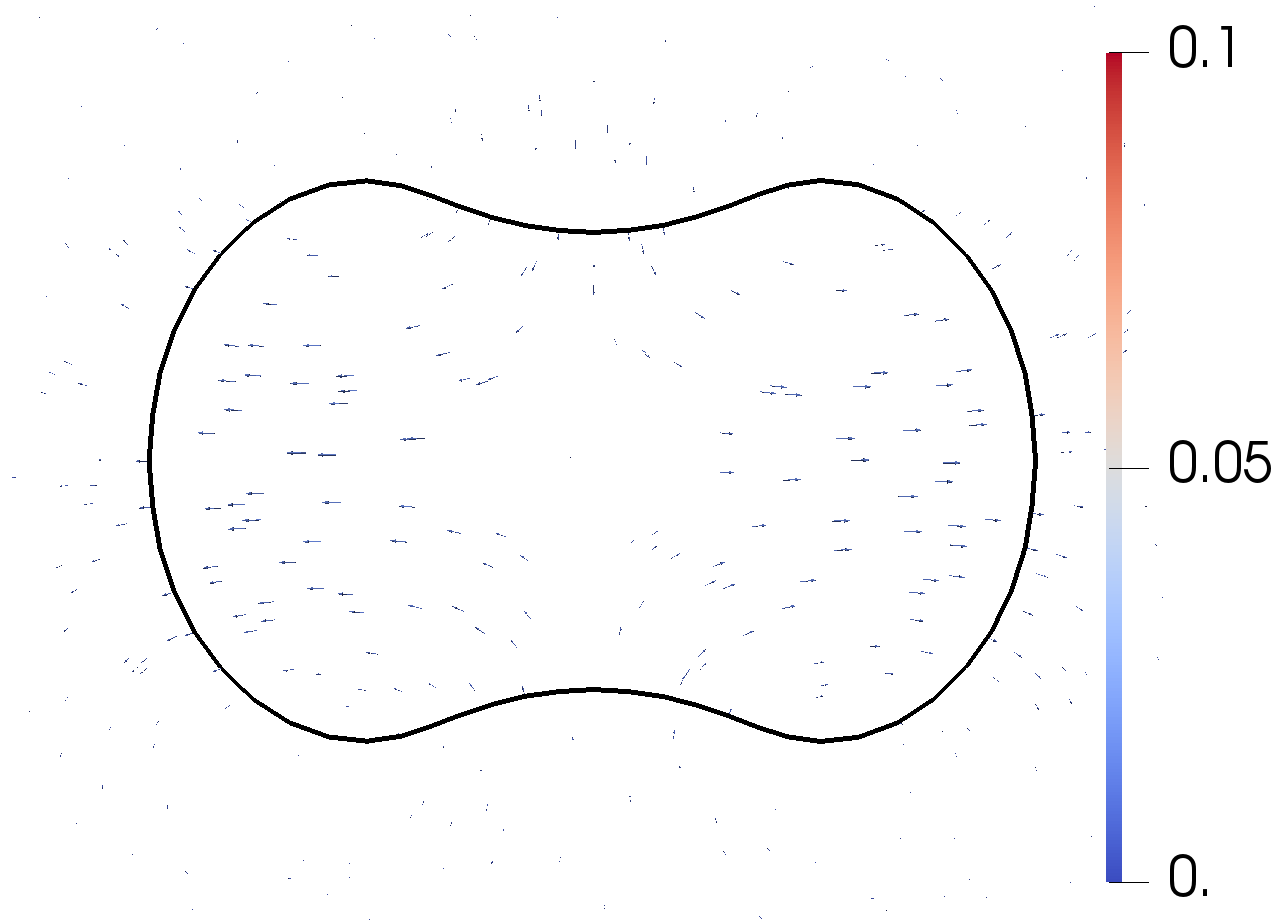}
        \caption{$t=14$}
        \label{fig:WOmarangoniV37}
    \end{subfigure}
    \caption{Spherical vesicle during the onset of cytokinesis, with surface tension force, $\gamma\nabla_\Gamma \cdot P$, i.e. without Marangoni term. 
    In (a)-(c) the colour scaling represents the norm of the force generated by the surface tension.
    (d)-(f) shows the cross-section in the $x$,$y$-plane with cell surface in black and velocity vectors colored by their magnitude. 
    All parameter values are chosen equal to 1.}
    \label{fig:WOmarangoni}
\end{figure}

We illustrate the deformation and fluid flows in Figures \ref{fig:cellDivSurfTen} - \ref{fig:WOmarangoni}. The form of surface tension force used in Figure \ref{fig:cellDivSurfTen} is $\nabla_\Gamma \cdot (\gamma P)$. This definition intrinsically includes the active Marangoni force contribution which drives surface flow along the gradient of surface tension (i.e. the force $\nabla_\Gamma \gamma$). This is seen by the strong, almost parallel, flow towards the equator. 

The surface stress triggers deformation of the membrane which in turn triggers fluid flow inside the cell, see Figures \ref{fig:CellDivV2}-\ref{fig:CellDivV37}. We observe that the velocity is highest where the norm of the surface tension force is highest and flowing parallel to the vesicle. The flow field induced by the surface tension results in two vortices with their centre located within the  vesicle and rotated around the $z$-axis.

The flow leads to a contraction of the equator along the ring of elevated surface tension. This contraction stops around $t=3.5$ and the cell assumes a stationary shape. In this state, steady flows are maintained within the cell, while flow at the cell surface is  purely tangential.

In Figure \ref{fig:WOmarangoni}, the surface tension force is defined as $\gamma (\nabla_\Gamma \cdot P)$. This essentially removes the Marangoni force contribution, resulting in a force perpendicular to the surface. We again observe two vortices in a ring around the vesicle. However, the flow is not mostly parallel to the surface as was the case with the Marangoni force contribution, but more perpendicular. This results in a more pronounced inward flow field at the equator, which leads to formation of a contractile ring with a neck. 
Having a purely normal surface force implies that the velocity field vanishes in the stationary state and the final shape resembles a dumbbell in which mean curvature times $\gamma$ is constant all over the surface. 
We conclude that the presence of Marangoni flow interestingly suppresses neck formation. 
This behavior might be due to the lack of the positive feedback mechanism occurring in active surface tension as described in \cite{Mietke2018}. 

The present model provides the first simulation of the onset of cytokinesis using viscoelastic surface rheology. 
In the future we will introduce the positive feedback mechanism by coupling the system to the transport of the force-generating molecules along the surface.
This will permit to investigate the influence of viscoelastic surface rheology on the observed pattern formation and shape dynamics and to tackle some of the open questions on cytokinesis \cite{10.1083/jcb.201612068}.

\section{Conclusion}
\label{sec:conclusion}
In this paper, we present the first numerical method for simulating the dynamics of a freely deforming viscoelastic surface. 
To this end, we constructed a novel mathematical model for viscoelastic surface stresses which is the surface equivalent of the upper-convected Maxwell model. The model fixes inconsistencies of earlier models \cite{sagis2010modelling,sagis2011dynamic}. The separate handling of dilational and deviatoric stress components makes it possible to choose large ratios of dilational and shear surface parameters.
Surface stress is coupled to bulk hydrodynamics modeled by the Navier-Stokes equations of the surrounding fluids. 
We solve this mathematical model numerically using an ALE-method with a Finite Element discretization with special Taylor-Hood elements.

Numerical results indicate good agreement with analytical solutions for two simplified test cases. In particular we showed that if the surface is flat, the numerical result converges to the analytical prediction of a two-dimensional Maxwell fluid as the viscosity of surrounding fluids goes to zero.
We illustrate the potential of the method by providing the first simulation of a viscoelastic fluid surface in shear flow. Also, we provide a phase diagram showing how viscoelastic parameters determine the transition from tank-treading to tumbling behavior. 
These results can be extended in a more detailed follow-up study. 

Finally, we present the first exemplary simulation of cytokinesis under a viscoelastic surface rheology, which illustrates the capability to use the method to tackle some of the open questions on cytokinesis \cite{10.1083/jcb.201612068} in the future. This will be addressed in an axisymmetric follow-up study.

\vspace{0.5cm}
\noindent
{\bf ACKNOWLEDGEMENTS} 
SA acknowledges support from the German Research Foundation DFG (grant AL1705/6) from DFG Research Unit FOR-3013 and 
support from the Saxon Ministry for Science and Art (SMWK MatEnUm-2).
Simulations were performed at the Center for Information Services and High Performance Computing (ZIH) at TU Dresden.

\section*{Appendix: Inclination angle and tank-treading frequency}
The inclination angle $\alpha$ is calculated using the moment of area tensor defined as in the appendix of \cite{Mokbel2017}. The eigenvectors of the moment of area tensor give the directions of the radii of the ellipse. Hence $\alpha$ is the angle between the $x$-axis and the eigenvector belonging to the largest eigenvalue.\\
\\
The tank-treading frequency is only defined when $\alpha$ is constant. $\omega$ is calculated by taking the average tank-treading frequency of each grid point of the mesh. The tank-treading frequency of a grid point with velocity $\vv{v}$ and location $\vv{r}$ is 
\[
\frac{||\vv{v} \times (\vv{r}-\vv{r_0})||_2}{||(\vv{r}-\vv{r_0})||_2^2},
\]
with $\vv{r_0}$ being the centre of the ellipsoid. 


\printbibliography

\end{document}